\documentclass[11pt]{article}
\usepackage{cite}
\usepackage[numbers,sort&compress]{natbib}

\usepackage{amsmath}
\usepackage{graphicx}
\usepackage{amsfonts}
\usepackage{bm}
\usepackage{amssymb}
\usepackage{epsfig}
\usepackage{color}
\usepackage{psfrag}
\usepackage{mathrsfs}
\usepackage{esint}
\usepackage{pifont}
\usepackage{amsthm}
\usepackage{url}

\usepackage{nicefrac} % For comparison
\numberwithin{equation}{section}
\usepackage{stackengine}
\usepackage{comment}
\usepackage{stmaryrd,scalerel} % questa package genera la doppia parentesi quadra: \llbracket \rrbracket % \llparenthesis 
\usepackage{subcaption}

\usepackage{tikz}
\usetikzlibrary{calc}
\usepackage{relsize}
\usepackage{bm}
\usepackage{moresize}
\usepgflibrary{shapes.geometric}
\usetikzlibrary{chains, positioning, quotes}
\usetikzlibrary{shapes,snakes}
\usetikzlibrary{snakes}
\usetikzlibrary{automata}
\usetikzlibrary{positioning}
\usetikzlibrary{arrows,decorations.pathmorphing,patterns,backgrounds,positioning,fit,petri}
\usetikzlibrary{automata}
\usetikzlibrary{graphs}
\usetikzlibrary{shadows}
\definecolor{colordiagram1}{RGB}{155, 255, 217}
\definecolor{colordiagram2}{RGB}{255, 148, 37} %colordiagram2
\definecolor{myyellow}{RGB}{255, 207, 0}
\newcommand{\Os}{ {\color{black}{ {\mathcal O} }} }

\def\dashint{\,\ThisStyle{\ensurestackMath{\stackinset{c}{.2\LMpt}{c}{.5\LMpt}{\SavedStyle-}{\SavedStyle\phantom{\int}}}\setbox0=\hbox{$\SavedStyle\int\,$}\kern-\wd0}\int}
\begin{document} \setcounter{page}{0}
\topmargin 0pt
\oddsidemargin 5mm
\renewcommand{\thefootnote}{\arabic{footnote}}
\newpage
\setcounter{page}{1}
\topmargin 0pt
\oddsidemargin 5mm
\renewcommand{\thefootnote}{\arabic{footnote}}
\newpage

\begin{titlepage}
\begin{flushright}
%SISSA 40/2012/EP \\
%DFTT 9/2007
\end{flushright}
\vspace{0.5cm}
\begin{center}
{\large {\bf Four-point interfacial correlation functions in two dimensions.\\Exact results from field theory and numerical simulations}}\\
\vspace{1.8cm}
{\large Alessio Squarcini$^{1,2,\natural}$ and Antonio Tinti$^{3,\flat}$ }\\
\vspace{0.5cm}
{\em $^1$Max-Planck-Institut f\"ur Intelligente Systeme,\\
Heisenbergstr. 3, D-70569, Stuttgart, Germany}\\
{\em $^2$IV. Institut f\"ur Theoretische Physik, Universit\"at Stuttgart,\\
Pfaffenwaldring 57, D-70569 
Stuttgart, Germany}\\
{\em $^3$Dipartimento di Ingegneria Meccanica e Aerospaziale,\\
Sapienza Universit\`a di Roma, via Eudossiana 18, 00184 Rome, Italy }\\
\end{center}
\vspace{1.2cm}

\renewcommand{\thefootnote}{\arabic{footnote}}
\setcounter{footnote}{0}

\begin{center}
\today
\end{center}

\begin{abstract}
We derive exact analytic results for several four-point correlation functions for statistical models exhibiting phase separation in two-dimensions. Our theoretical results are then specialized to the Ising model on the two-dimensional strip and found to be in excellent agreement with high-precision Monte Carlo simulations.
\end{abstract}

\vfill
$^\natural$squarcio@is.mpg.de, $^\flat$antonio.tinti@uniroma1.it
\end{titlepage}

%\newpage
%\tableofcontents
%\newpage

%=====================================================================================
\section{Introduction}
\label{sec1}
The study of extended objects subjected to thermal fluctuations such as interfaces, membranes, and polymers, constitutes a fundamental chapter in statistical physics \cite{NPW}. Interfaces separating coexisting phases, as in the case of a liquid in equilibrium with its vapor \cite{Evans_79}, have received considerable attention from both theory, simulations \cite{BLM}, and experiments \cite{ASL_2004}; we refer to \cite{Widom_1972, Jasnow_1984, gennes_wetting_1985, sullivantelodagama, diehl_fieldtheoretical_1986, dietrich_wetting_1988, MSchick, FLN, BEIMR} for reviews on this subject.

Within the language of statistical physics, the characterization of thermal fluctuations exhibited by extended objects is inevitably formulated in terms of correlation functions of certain quantities. This is also the case for liquid-vapor interfaces. In a seminal paper by Wertheim \cite{Wertheim_1976} it was shown that density fluctuations in the presence of phase separation become long-ranged along the interface separating coexisting phases and that these correlations are confined within the interfacial region \cite{Weeks_1977, Evans_79}. This situation happens to be in sharp contrast with the exponential decay of correlations which characterizes pure phases\footnote{See \cite{AK1977,CIV2003} for rigorous results in the framework of Ornstein-Zernike theory.} \cite{Stanley, hansen_theory_2006}. The above features exhibited by correlations within interfacial region separating coexisting phases have been at the center of numerous studies based on coarse-grained descriptions in terms of effective interfacial models \cite{BW_1985, MD_1999, PRWE_2014, HD_2015, CT_2016} followed from the so-called capillary wave model \cite{BLS_1965}. We refer to \cite{RowlinsonWidom} for a historical account and to \cite{PRWE_2014,HD_2015,CT_2016} (and references therein) for the state of the art on effective interfacial models.

Going beyond effective models, and focusing on exactly solvable models, for long time the study of interfacial behavior has been limited to the planar Ising model on the lattice \cite{Abraham_review}. The possibility of obtaining analytic results from the scaling limit of exact solutions on the lattice proved to be essential for the formulation of heuristic interpretations based on the analogy between fluctuating interfaces in two dimensions and random walks \cite{fisher_walks_1984} as well as formulations based on Solid-On-Solid models \cite{Kroll,Burkhardt1981} in which spin interfaces are identified with fluctuating Onsager-Temperley strings \cite{temperley_1952} and statistical sums are implemented by path integrals \cite{VL, Burkhardt}.

The two-dimensional case turns out to be very interesting because the exact analytic form of interfacial correlations can be obtained for several models of statistical mechanics \cite{DS_twopoint}. In particular, it is possible to characterize in a mathematically precise form the anisotropic character of interfacial correlations, their long-range character, as well as their confinement, providing thus a quantitative analysis of the scenario outlined by Wertheim within the context of non uniform fluids \cite{Wertheim_1976,Weeks_1977}.

In this paper, we show how to find closed-form expressions for certain four-point correlation functions of the order parameter field in the presence of phase separation. The systems considered in this paper are generic models of two-dimensional statistical mechanics at phase coexistence close to a second-order phase transition point. The analytic expressions derived in this paper follow from the exact theory of phase separation in two dimensions\footnote{The three-dimensional case has been recently addressed in \cite{DSS_2020,DSS_2021}.} \cite{DV, DS_bubbles, DS_twopoint} and wetting phenomena \cite{DS_wetting, DS_wedge, DS_wedgebubble, ST_droplet} and, more specifically, to recently obtained results for $n$-point correlation functions in the presence of an interface \cite{Squarcini_Multipoint}. In particular, it has been shown in \cite{Squarcini_Multipoint} how the first-principle field-theoretic calculation for $n$-point correlation functions admits an \emph{exact} probabilistic interpretation in which the interface fluctuates as a Brownian bridge. The results obtained in this paper are thus self-contained and derived by following the probabilistic interpretation, which is exact at both leading and first subleading corrections in finite-size corrections.

This paper is structured as follows. In Sec.~\ref{sec2}, we recall the probabilistic interpretation and we introduce the specific four-point correlation functions considered throughout this paper. We then derive the corresponding analytic expression for each order parameter correlator and discuss some analytic properties they satisfy. In Sec.~\ref{sec3}, we present the comparison between the analytic expressions and the results obtained from Monte Carlo simulations for the Ising model. Conclusive remarks are summarized in Sec.~\ref{sec4}. Appendix \ref{Appendix_A} collects some mathematical details involved in the calculations presented in Sec.~\ref{sec2}.

%=====================================================================================
\section{Interfacial correlations}
\label{sec2}
\begin{figure*}[htbp]
\centering
%\hspace{8mm}
        \begin{subfigure}[b]{0.3\textwidth}
            \centering
\begin{tikzpicture}[thick, line cap=round, >=latex, scale=0.24]
\tikzset{fontscale/.style = {font=\relsize{#1}}}
\def\L{9}
\def\y{6}
\draw[thin, dashed, -] (-10, 0) -- (11, 0) node[below] {};
\draw[thin, dashed, -] (0, - \L) -- (0, \L) node[left] {};
\draw[very thick, red, -] (0, \L) -- (10, \L) node[] {};
\draw[very thick, red, -] (0, -\L) -- (10, -\L) node[] {};
\draw[very thick, blue, -] (-10, \L) -- (0, \L) node[] {};
\draw[very thick, blue, -] (-10, -\L) -- (0, -\L) node[] {};
\draw[thin, fill=green!30] (0, 7) circle (10pt) node[left] { $(0,y_{1})$ };;
\draw[thin, fill=green!30] (0, 3) circle (10pt) node[left] { $(0,y_{2})$ };;
\draw[thin, fill=green!30] (0, -1) circle (10pt) node[left] { $(0,y_{3})$ };;
\draw[thin, fill=green!30] (0, -6) circle (10pt) node[left] { $(0,y_{4})$ };;
\draw[thin, fill=white] (-5, \L) circle (0pt) node[above] {${\color{blue}{a}}$};;
\draw[thin, fill=white] (-5, -\L) circle (0pt) node[below] {${\color{blue}{a}}$};;
\draw[thin, fill=white] (5, \L) circle (0pt) node[above] {${\color{red}{b}}$};;
\draw[thin, fill=white] (5, -\L) circle (0pt) node[below] {${\color{red}{b}}$};;
\end{tikzpicture}
            \caption[]%
            {{\small $G_{\parallel}(y_{1},y_{2},y_{3},y_{4})$}}    
            \label{fig_ssss_a}
        \end{subfigure}
\hfill
        \begin{subfigure}[b]{0.3\textwidth}
            \centering
\begin{tikzpicture}[thick, line cap=round, >=latex, scale=0.24]
\tikzset{fontscale/.style = {font=\relsize{#1}}}
\def\L{9}
\def\y{6}
\draw[thin, dashed, -] (-10, 0) -- (11, 0) node[below] {};
\draw[thin, dashed, -] (0, - \L) -- (0, \L) node[left] {};
\draw[very thick, red, -] (0, \L) -- (10, \L) node[] {};
\draw[very thick, red, -] (0, -\L) -- (10, -\L) node[] {};
\draw[very thick, blue, -] (-10, \L) -- (0, \L) node[] {};
\draw[very thick, blue, -] (-10, -\L) -- (0, -\L) node[] {};
\draw[thin, fill=green!30] (0, 6) circle (10pt) node[left] { $(0,y)$ };;
\draw[thin, fill=green!30] (0, -6) circle (10pt) node[left] { $(0,-y)$ };;
\draw[thin, fill=green!30] (7, 0) circle (10pt) node[below] { $(x,0)$ };;
\draw[thin, fill=green!30] (-7, 0) circle (10pt) node[below] { $(-x,0)$ };;
\draw[thin, fill=white] (-5, \L) circle (0pt) node[above] {${\color{blue}{a}}$};;
\draw[thin, fill=white] (-5, -\L) circle (0pt) node[below] {${\color{blue}{a}}$};;
\draw[thin, fill=white] (5, \L) circle (0pt) node[above] {${\color{red}{b}}$};;
\draw[thin, fill=white] (5, -\L) circle (0pt) node[below] {${\color{red}{b}}$};;
\end{tikzpicture}
            \caption[]%
            {{\small $G_{\diamond}(x,y)$}}    
            \label{fig_ssss_b}
        \end{subfigure}
\hfill
        \begin{subfigure}[b]{0.3\textwidth}
            \centering
\begin{tikzpicture}[thick, line cap=round, >=latex, scale=0.24]
\tikzset{fontscale/.style = {font=\relsize{#1}}}
\def\L{9}
\def\y{6}
\draw[thin, dashed, -] (-10, 0) -- (11, 0) node[below] {};
\draw[thin, dashed, -] (0, - \L) -- (0, \L) node[left] {};
\draw[very thick, red, -] (0, \L) -- (10, \L) node[] {};
\draw[very thick, red, -] (0, -\L) -- (10, -\L) node[] {};
\draw[very thick, blue, -] (-10, \L) -- (0, \L) node[] {};
\draw[very thick, blue, -] (-10, -\L) -- (0, -\L) node[] {};
\draw[thin, fill=green!30] (5, \y) circle (10pt) node[below] { $(x,y)$ };;
\draw[thin, fill=green!30] (5, -\y) circle (10pt) node[above] { $(x,-y)$ };;
\draw[thin, fill=green!30] (-5, \y) circle (10pt) node[below] { $(-x,y)$ };;
\draw[thin, fill=green!30] (-5, -\y) circle (10pt) node[above] { $(-x,-y)$ };;
\draw[thin, fill=white] (-5, \L) circle (0pt) node[above] {${\color{blue}{a}}$};;
\draw[thin, fill=white] (-5, -\L) circle (0pt) node[below] {${\color{blue}{a}}$};;
\draw[thin, fill=white] (5, \L) circle (0pt) node[above] {${\color{red}{b}}$};;
\draw[thin, fill=white] (5, -\L) circle (0pt) node[below] {${\color{red}{b}}$};;
\end{tikzpicture}
            \caption[]%
            {{\small $G_{\oblong}(x,y)$}}    
            \label{fig_ssss_c}
        \end{subfigure}
%\hspace{8mm}
\vskip\baselineskip
\caption[]
{\small Spatial arrangement of the order parameter fields (green circles) for the four-point correlation functions considered in this paper.}
\label{fig_ssss}
\end{figure*}
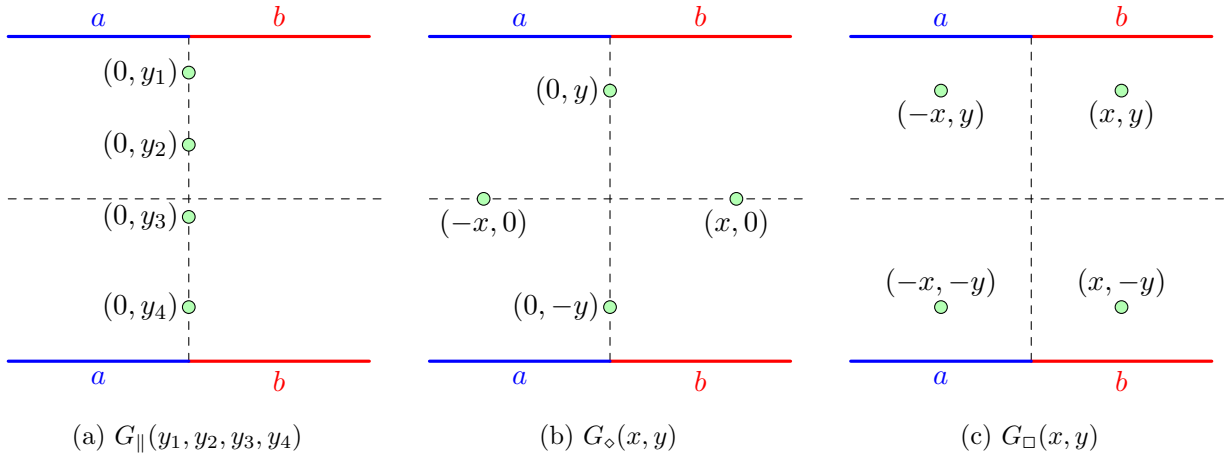

We consider a two-dimensional statistical system at phase coexistence close to a second order phase transition point. To be definite, the system we consider is a ferromagnetic spin model defined on the two-dimensional strip $(x,y) \in \mathbb{R} \times (-R/2,R/2)$ with finite width $R$ in the $y$ direction. Boundary conditions with spins fixed in state $a$ for $x<0$ and $b$ for $x>0$ are used to enforce the emergence of an interface separating coexisting phases; see Fig.~\ref{fig_ssss}.

The quantity which we are going to examine is the four-point correlation function of the order parameter field $\sigma(x,y)$
\begin{equation}
\label{001}
\langle \sigma(x_{1},y_{1}) \sigma(x_{2},y_{2}) \sigma(x_{3},y_{3}) \sigma(x_{4},y_{4}) \rangle_{ab} \, ,
\end{equation}
where $\langle \cdots \rangle_{ab}$ stands for statistical averages in the strip geometry with $ab$ boundary conditions corresponding to Fig.~\ref{fig_ssss}. It has been shown in Ref.~\cite{Squarcini_Multipoint} how general results for $n$-point interfacial correlation functions with arbitrary $n$ can be derived from the underlying field theory associated to the scaling model. In this paper, we specify the above mentioned results to $n=4$ with the four points arranged in particularly symmetric configurations. This choice will allow us to write the exact results of Ref.~\cite{Squarcini_Multipoint} in an explicit form which is particularly suited for both analytical considerations as well as for comparison with numerical simulations.

The configurations we will examine throughout this paper are those depicted in Fig.~\ref{fig_ssss}, and the corresponding four-point correlation functions are those defined in (\ref{11022021_1325})-(\ref{11032021_0904}). The correlation function
\begin{equation}
\label{11022021_1325}
G_{\parallel}(y_{1},y_{2},y_{3},y_{4}) = \langle \sigma(0,y_{1}) \sigma(0,y_{2}) \sigma(0,y_{3}) \sigma(0,y_{4}) \rangle_{ab}
\end{equation}
corresponds to order parameter fields aligned along the straight line which connects the pinning points, as illustrated in Fig.~\ref{fig_ssss_a}. The correlation function
\begin{equation}
\label{11032021_0903}
G_{\diamond}(x,y) = \langle \sigma(0,y) \sigma(-x,0) \sigma(x,0) \sigma(0,-y) \rangle_{ab}
\end{equation}
corresponds to order parameter fields arranged in a rhomboidal pattern with diagonals of length $2x$ and $2y$ parallel to the coordinate axes, as indicated in Fig.~\ref{fig_ssss_b}. Lastly, the correlation function
\begin{eqnarray}
\label{11032021_0904}
G_{\oblong}(x,y) = \langle \sigma(-x,y) \sigma(x,y) \sigma(-x,-y) \sigma(x,-y) \rangle_{ab}
\end{eqnarray}
corresponds to order parameter fields arranged in a rectangular pattern with edges of length $2x$ and $2y$ parallel to the coordinate axes, as depicted in Fig.~\ref{fig_ssss_c}. Throughout this manuscript we will refer to $G_{\parallel}(y_{1},y_{2},y_{3},y_{4})$, $G_{\diamond}(x,y)$ and $G_{\oblong}(x,y)$ as the \emph{parallel}, \emph{rhomboidal} and \emph{rectangular} correlation functions, respectively.

We are now in the position to introduce the theoretical framework employed to calculate the correlation functions we are interested in. To this end, we recall those results obtained in \cite{Squarcini_Multipoint} which are essential for the discussion in the present paper. It has been shown in \cite{Squarcini_Multipoint} that exact results for $n$-point correlation function -- such as (\ref{001}) when $n=4$ -- can be represented within a probabilistic interpretation in which the correlation function is reconstructed by averaging over interfacial configurations weighted with the probability density of a Brownian bridge. This probabilistic interpretation is valid for those universality classes and boundary conditions in which phase separation in two dimensions occurs through a single interface \cite{DV,DS_bubbles,Delfino_localization}.

More specifically, the probabilistic interpretation allows us to express the correlation function (\ref{001}) as follows
\begin{equation}
\begin{aligned}
\label{a004a}
\langle \sigma_{1}(x_{1},y_{1}) \sigma_{2}(x_{2},y_{2}) \sigma_{3}(x_{3},y_{3}) \sigma_{4}(x_{4},y_{4}) \rangle_{ab} & = \int_{\mathbb{R}^{4}}\textrm{d}u_{1} \textrm{d}u_{2} \textrm{d}u_{3} \textrm{d}u_{4} \, P_{4}(u_{1},y_{1}; u_{2},y_{2}; u_{3},y_{3}; u_{4},y_{4}) \times \\
& \times \sigma_{ab\vert 1}(x_{1} \vert u_{1}) \sigma_{ab \vert 2}(x_{2} \vert u_{2}) \sigma_{ab \vert 3}(x_{3} \vert u_{3}) \sigma_{ab \vert 4}(x_{4} \vert u_{4}) \, .
\end{aligned}
\end{equation}
Some comments are in order. The subscript $j$ in $\sigma_{j}$ denotes the $j$-th spin field entering in (\ref{a004a}). In the more general setting, the spin field $\sigma_{j}$ can have more than one component and $j$ can be used to label them. For instance, this situation happens in the quantum field theory associated to the scaling $q$-state Potts model \cite{DV}. Then, $P_{4}\textrm{d}u_{1} \textrm{d}u_{2} \textrm{d}u_{3} \textrm{d}u_{4}$ is the probability for the interface to cross \emph{all} the intervals $(x_{j},x_{j}+\textrm{d}x_{j})$ at $y=y_{j}$ with $j=1,2,3$ and $4$. As shown in \cite{Squarcini_Multipoint}, the joint passage probability $P_{4}$ is the one which characterizes a Brownian bridge in which the random walker starts its motion in $x=0$ at time $y=-R/2$ and comes back to $x=0$ at time $y=R/2$ (the specific form of $P_{n}$ is provided in (\ref{11032021_1534})). The profile
\begin{equation}
\label{a005a}
\sigma_{ab \vert i}(x \vert u) = \langle \sigma_{i} \rangle_{a} \theta(u-x) + \langle \sigma_{i} \rangle_{b} \theta(x-u) + A_{ab \vert i}^{(\sigma_{i})} \delta(x-u) + \dots
\end{equation}
gives the magnetization at point $(x,y)$ when the interface is conditioned to cross the point $(u,y)$. In (\ref{a005a}), $\theta(x)$ denotes Heaviside unit step function, i.e., $\theta(x)=1$ if $x>0$ and $\theta(x)=0$ if $x<0$, and $\delta$ stands for Dirac delta function. 

The first two terms in the right hand side of (\ref{a005a}) correspond to a step profile in which the two coexisting phases are sharply separated by the interface. According to this picture, the interface sharply separates two regions in which the order parameter exhibits the vacuum expectation values $\langle \sigma_{i} \rangle_{a}$ and $\langle \sigma_{i} \rangle_{b}$. The subsequent term proportional to the Dirac delta takes into account the first correction beyond the picture in which the interface is regarded as a structureless entity. The interface structure term originates a subdominant correction which contributes at order $R^{-1/2}$ and whose overall factor $A_{ab \vert i}^{(\sigma_{i})}$ is known for integrable field theories \cite{DV}; we refer to \cite{DSS1} for recent numerical simulations. We also mention that subsequent corrections can be systematized in an exact expansion in powers of the small parameter $\sqrt{\xi/R}$ and that corrections at order $R^{-\ell/2}$ with $\ell=2,3,\dots$ can be obtained within the field-theoretical method outlined in \cite{Squarcini_Multipoint}. Since in this paper we are interested in the leading-order form of interfacial correlations, from now on we retain the first two addends in (\ref{a005a}). 

Let us discuss the general structure of the leading-order result for the correlation function. It is straightforward to realize how the Heaviside theta functions in (\ref{a004a}) yield a representation in which the correlation function is expressed through cumulative distributions of the probability density $P_{4}$. It turns out to be convenient to write the first two terms in the right hand side of (\ref{a005a}) in the form
\begin{equation}
\label{ }
\langle \sigma_{i} \rangle_{a} - \Delta \langle \sigma_{i} \rangle \theta(x-u) \, ,
\end{equation}
with the jump of expectation values in pure phases given by $\Delta \langle \sigma_{i} \rangle=\langle \sigma_{i} \rangle_{a} - \langle \sigma_{i} \rangle_{b}$. The joint passage probability $P_{4}$ satisfies the following properties:
\begin{align}
\label{}
P_{3}(u_{1},y_{1}; u_{2},y_{2}; u_{3},y_{3}) & = \int_{\mathbb{R}}\textrm{d}u_{4} \, P_{4}(u_{1},y_{1}; u_{2},y_{2}; u_{3},y_{3}; u_{4},y_{4}) \\
P_{2}(u_{1},y_{1}; u_{2},y_{2}) & = \int_{\mathbb{R}^{2}}\textrm{d}u_{3}\textrm{d}u_{4} \, P_{4}(u_{1},y_{1}; u_{2},y_{2}; u_{3},y_{3}; u_{4},y_{4}) \\
P_{1}(u_{1},y_{1}) & = \int_{\mathbb{R}^{3}}\textrm{d}u_{2}\textrm{d}u_{3}\textrm{d}u_{4} \, P_{4}(u_{1},y_{1}; u_{2},y_{2}; u_{3},y_{3}; u_{4},y_{4}) \\
1 & = \int_{\mathbb{R}^{4}}\textrm{d}u_{1}\textrm{d}u_{2}\textrm{d}u_{3}\textrm{d}u_{4} \, P_{4}(u_{1},y_{1}; u_{2},y_{2}; u_{3},y_{3}; u_{4},y_{4}) \, ;
\end{align}
the first three equations give the marginal distributions and the last one is the normalization condition. Analogously to $P_{4}$, the joint probabilities $P_{n}$ express the probability density for the crossing of $n$ intervals. It has been shown in \cite{Squarcini_Multipoint} that that $P_{n}$ can be expressed in terms of the $n$-variate normal distribution $\Pi_{n}$ \cite{AG,Tong} through
\begin{equation}
\label{11032021_1534}
P_{n}(x_{1},y_{1}; \cdots; x_{n},y_{n}) = \left( \prod_{j=1}^{n} \frac{\sqrt{2}}{\kappa_{j}\lambda} \right) \Pi_{n}(\sqrt{2}\chi_{1},\dots,\sqrt{2}\chi_{n} \vert \textsf{R}_{1\dots n}) \, ,
\end{equation}
with the following notations
\begin{equation}
\label{ }
\chi_{j} = x_{j}/(\kappa_{j}\lambda) \, , \quad \kappa_{j} = \sqrt{1-\tau_{j}^{2}} \, \quad , \quad \tau_{j} = 2y_{j}/R \, , \quad \lambda = \sqrt{R/(2m)} \, ;
\end{equation}
the parameter $m$, which is the kink mass in field theory, and the surface tension in the language of liquid state theories \cite{DV}, is related to the subcritical bulk correlation length $\xi$ via $\xi=1/(2m)$. Then, $\Pi_{n}(u_{1},\dots,u_{n} \vert \textsf{R}_{1\dots n})$ is the $n$-variate normal distribution with (symmetric) correlation matrix $\textsf{R}_{1\dots n}$. The $ij$ matrix element of the correlation matrix $\rho_{ij}=(\textsf{R}_{1\dots n})_{ij}$ is given by
\begin{equation}
\label{04042021_2028}
\rho_{ij} = \sqrt{ \frac{1-\tau_{i}}{1+\tau_{i}} \frac{1+\tau_{j}}{1-\tau_{j}} } \, \quad i \leqslant j \, .
\end{equation}
The elements in the lower triangular part follow by recalling that $\textsf{R}_{1\dots n}$ is a symmetric matrix. Note that we are taking $y_{i}>y_{j}$ for $i<j$, and so also $\tau_{i}>\tau_{j}$ for $i<j$. Expectation values with measure $\Pi_{n}$ are defined by
\begin{equation}
\label{ }
\mathbb{E}[ f(u_{1},\dots,u_{n}) ] = \int_{\mathbb{R}^{n}}\textrm{d}u_{1}\dots\textrm{d}u_{n} \, f(u_{1},\dots,u_{n}) \Pi_{n}(u_{1},\dots,u_{n} \vert \textsf{R}_{1,\dots,n}) \, .
\end{equation}
The probability density is standardized, meaning that $\mathbb{E}[ u_{i}^{2} ] = 1$ and 
\begin{equation}
\label{ }
\mathbb{E}[ u_{i}u_{j} ] = \rho_{ij} \, .
\end{equation}
The normalization condition then ensures that $\mathbb{E}[ 1 ] = 1$. The last quantity we need to introduce is the cumulative function of the $n$-variate normal distribution
\begin{equation}
\label{ }
F_{n}(X_{1},\dots,X_{n} \vert \textsf{R}_{1\dots n}) = \int_{-\infty}^{X_{1}}\textrm{d}u_{1} \dots \int_{-\infty}^{X_{n}}\textrm{d}u_{n} \, \Pi_{n}(u_{1},\dots,u_{n} \vert \textsf{R}_{1\dots n}) \, .
\end{equation}
The above allows us to cast integrals arising from the probabilistic interpretation in terms of cumulative functions. For instance
\begin{equation}
\label{ }
\int_{-\infty}^{x_{1}}\textrm{d}u_{1} \cdots \int_{-\infty}^{x_{4}}\textrm{d}u_{4} \, P_{4}(u_{1},y_{1}; u_{2},y_{2}; u_{3},y_{3}; u_{4},y_{4}) = F_{4}(\sqrt{2}\chi_{1},\dots,\sqrt{2}\chi_{4} \vert \textsf{R}_{1234}) \, .
\end{equation}
By proceeding along the above lines, the four-point correlation function (\ref{001}) reads
\begin{equation}
\begin{aligned}
\label{a006a}
\langle \sigma_{i}(x_{1},y_{1}) \sigma_{i}(x_{2},y_{2}) \sigma_{i}(x_{3},y_{3}) \sigma_{i}(x_{4},y_{4}) \rangle_{ab} & = \langle \sigma_{i} \rangle_{a}^{4} - \langle \sigma_{i} \rangle_{a}^{3} \Delta \langle \sigma_{i} \rangle \sum_{j=1}^{4} F_{1}(\sqrt{2}\chi_{j}) \\
& + \langle \sigma_{i} \rangle_{a}^{2} (\Delta \langle \sigma_{i} \rangle)^{2} \sum_{1 \leqslant i<j \leqslant 4} F_{2}(\sqrt{2}\chi_{i},\sqrt{2}\chi_{j} \vert \textsf{R}_{ij}) \\
& - \langle \sigma_{i} \rangle_{a} (\Delta \langle \sigma_{i} \rangle)^{3} \sum_{1 \leqslant i<j<k \leqslant 4} F_{3}(\sqrt{2}\chi_{i},\sqrt{2}\chi_{j},\sqrt{2}\chi_{k} \vert \textsf{R}_{ijk} ) \\
& + (\Delta \langle \sigma_{i} \rangle)^{4} F_{4}(\sqrt{2}\chi_{1},\sqrt{2}\chi_{2},\sqrt{2}\chi_{3},\sqrt{2}\chi_{4} \vert \textsf{R}_{1234} ) \\
& + \Os(R^{-1/2}) \, ,
\end{aligned}
\end{equation}
where all spin fields carry the same index $i$. The most general case in which different components of the order parameter occur in (\ref{a006a}) can be treated along the same lines. It is worth pointing out how clustering properties of correlation functions can be derived from (\ref{a006a}). For instance, by taking $x_{4}\rightarrow \mp \infty$, we have the following limiting behavior
\begin{equation}
\label{ }
\lim_{x_{4}\rightarrow \mp \infty} \langle \sigma_{i}(x_{1},y_{1}) \sigma_{i}(x_{2},y_{2}) \sigma_{i}(x_{3},y_{3}) \sigma_{i}(x_{4},y_{4}) \rangle_{ab} = \langle \sigma_{i} \rangle_{a [b]} \langle \sigma_{i}(x_{1},y_{1}) \sigma_{i}(x_{2},y_{2}) \sigma_{i}(x_{3},y_{3}) \rangle_{ab} \, ,
\end{equation}
where the three-point correlation function
\begin{equation}
\begin{aligned}
\langle \sigma_{i}(x_{1},y_{1}) \sigma_{i}(x_{2},y_{2}) \sigma_{i}(x_{3},y_{3}) \rangle_{ab} & = \langle \sigma_{i} \rangle_{a}^{3} - \langle \sigma_{i} \rangle_{a}^{2} \Delta \langle \sigma_{i} \rangle \sum_{j=1}^{3} F_{1}(\sqrt{2}\chi_{j}) \\
& + \langle \sigma_{i} \rangle_{a} (\Delta \langle \sigma_{i} \rangle)^{2} \sum_{1 \leqslant i<j \leqslant 3} F_{2}(\sqrt{2}\chi_{i},\sqrt{2}\chi_{j} \vert \textsf{R}_{ij}) \\
& - (\Delta \langle \sigma_{i} \rangle)^{3} F_{3}(\sqrt{2}\chi_{1},\sqrt{2}\chi_{2},\sqrt{2}\chi_{3} \vert \textsf{R}_{123} ) + \Os(R^{-1/2}) \, ,
\end{aligned}
\end{equation}
agrees with field-theoretic calculations \cite{Squarcini_Multipoint, ST_threepoint}.

The remaining part of this section is devoted to a detailed case-by-case analysis of (\ref{a006a}) with spin fields arranged as shown in Fig.~\ref{fig_ssss}. Such an analysis will allow us to find explicit expressions for (\ref{a006a}) involving either single integrals or certain special functions related to the Gaussian distribution \cite{Owen1956,Owen1980}.

%=====================================================================================
\subsection{Parallel correlation function}
\label{section31}
For spin fields arranged in the configuration of Fig.~\ref{fig_ssss_a}, we have $x_{j}=0$ with $j=1,\dots,4$ and $-R/2 < y_{4} < y_{3} < y_{2} < y_{1} < R/2$. Since $\chi_{j}=0$ in (\ref{a006a}), the calculation of the parallel correlation function involves \emph{orthant probabilities} only. We introduce the following shorthand notation for orthant probabilities:
\begin{equation}
\begin{aligned}
\label{30122020_0952}
\Phi_{2}(\rho_{12}) & = F_{2}(0,0\vert \textsf{R}_{12}) \\
\Phi_{3}(\rho_{12}, \rho_{23}) & = F_{3}(0,0,0 \vert \textsf{R}_{123}) \\
\Phi_{4}(\rho_{12}, \rho_{23}, \rho_{34}) & = F_{4}(0,0,0,0 \vert \textsf{R}_{1234}) \, .
\end{aligned}
\end{equation}
Clearly, $F_{1}(0)=1/2$. We recall that $\textsf{R}_{1\dots n}$ is a $n\times n$ symmetric matrix with $1$ along the main diagonal, therefore it comprises $n(n-1)/2$ nontrivial entries in the upper triangle. Thanks to the Markov property\footnote{See e.g. \cite{Doob,McFadden}.}, $\rho_{ij}\rho_{jk}=\rho_{ik}$ for $i<j<k$. It thus follows that the number of independent entries is actually lowered to $n-1$. For this reason the orthant probability for the trivariate normal distribution depends on two independent variables and, analogously, the orthant probability for the quadrivariate normal distribution depends on three independent variables. Without loss of generality, we can take for the independent correlation coefficients those which appear in the superdiagonal entries, namely, $\rho_{i,i+1}$ for $i=1,\dots,n-1$. Thanks to (\ref{a006a}) and (\ref{30122020_0952}) the parallel correlation function reads
\begin{equation}
\begin{aligned}
\label{a006b}
G_{\parallel}(y_{1},y_{2},y_{3},y_{4}) & = \langle \sigma_{i} \rangle_{a}^{4} - 2\langle \sigma_{i} \rangle_{a}^{3} \Delta \langle \sigma_{i} \rangle + \langle \sigma_{i} \rangle_{a}^{2} (\Delta \langle \sigma_{i} \rangle)^{2} \sum_{1 \leqslant i<j \leqslant 4} \Phi_{2}(\rho_{ij}) \\
& - \langle \sigma_{i} \rangle_{a} (\Delta \langle \sigma_{i} \rangle)^{3} \biggl[ \Phi_{3}(\rho_{12},\rho_{23}) + \Phi_{3}(\rho_{12},\rho_{24}) + \Phi_{3}(\rho_{13},\rho_{34}) \\
& + \Phi_{3}(\rho_{23},\rho_{34}) \biggr] + (\Delta \langle \sigma_{i} \rangle)^{4} \Phi_{4}(\rho_{12},\rho_{23},\rho_{34}) + \Os(R^{-1/2}) \, .
\end{aligned}
\end{equation}
The orthant probabilities $\Phi_{j}$ for the bivariate ($j=2$) and trivariate ($j=3$) normal distributions are given by \cite{Tong, AG}
\begin{equation}
\label{a007a}
\Phi_{2}(\rho_{12}) = \frac{1}{4} + \frac{1}{2\pi} \sin^{-1}(\rho_{12}) \, ,
\end{equation}
and
\begin{equation}
\label{a008a}
\Phi_{3}(\rho_{12},\rho_{23}) = \frac{1}{8} + \frac{1}{4\pi} \Bigl[ \sin^{-1}(\rho_{12}) + \sin^{-1}(\rho_{13}) + \sin^{-1}(\rho_{23}) \Bigr] \, ,
\end{equation}
where $\rho_{13}=\rho_{12}\rho_{23}$ because of the Markov property which characterizes the Brownian bridge. The expression of the orthant probability for the quadrivariate normal distribution is considerably more involved and reads \cite{Cheng}
\begin{equation}
\begin{aligned}
\label{a009a}
\Phi_{4}(\rho_{12},\rho_{23},\rho_{34}) & = \frac{1}{16} + \frac{1}{8\pi} \sum_{1 \leqslant i<j \leqslant 4} \sin^{-1}(\rho_{ij}) + \frac{1}{4\pi^{2}} \sin^{-1}(\rho_{12}) \sin^{-1}(\rho_{34}) \\
& + \frac{1}{4\pi^{2}} \int_{0}^{\rho_{12}\rho_{23}\rho_{34}} \frac{\textrm{d}u}{\sqrt{1-u^{2}}} \sin^{-1}\left( u \sqrt{ \frac{1-\rho_{12}^{2}}{\rho_{12}^{2}-u^{2}} \frac{1-\rho_{34}^{2}}{\rho_{34}^{2}-u^{2}} } \right) \, .
\end{aligned}
\end{equation}
By plugging (\ref{a007a})-(\ref{a009a}) into (\ref{a006b}), and introducing
\begin{equation}
\label{ }
\widetilde{ \langle \sigma_{i} \rangle } = \frac{ \langle\sigma_{i}\rangle_{a} + \langle\sigma_{i}\rangle_{b} }{ 2 } \, ,
\end{equation}
we find
\begin{equation}
\begin{aligned}
\label{a010a}
G_{\parallel}(y_{1},y_{2},y_{3},y_{4}) & = \widetilde{ \langle \sigma_{i} \rangle } ^{4} + \frac{1}{4} \widetilde{ \langle \sigma_{i} \rangle }^{2} \left( \Delta \langle \sigma \rangle \right)^{2} \mathcal{C}_{\parallel}(y_{1},y_{2},y_{3},y_{4}) + \frac{1}{16} \left( \Delta \langle \sigma_{i} \rangle \right)^{4} \mathcal{G}_{\parallel}(y_{1},y_{2},y_{3},y_{4}) \\
& + \Os(R^{-1/2}) \, ,
\end{aligned}
\end{equation}
where $\mathcal{C}_{\parallel}$ and $\mathcal{G}_{\parallel}$ are the scaling functions
\begin{equation}
\label{12042021_0843}
\mathcal{C}_{\parallel}(y_{1},y_{2},y_{3},y_{4}) = \frac{2}{\pi} \sum_{1 \leqslant i<j \leqslant 4} \sin^{-1}(\rho_{ij}) \, ,
\end{equation}
and
\begin{equation}
\begin{aligned}
\label{a011a}
\mathcal{G}_{\parallel}(y_{1},y_{2},y_{3},y_{4}) & = \frac{4}{\pi^{2}} \sin^{-1}(\rho_{12}) \sin^{-1}(\rho_{34}) \\
& + \frac{4}{\pi^{2}} \int_{0}^{\rho_{12}\rho_{23}\rho_{34}} \frac{\textrm{d}u}{\sqrt{1-u^{2}}} \sin^{-1}\left( u \sqrt{ \frac{1-\rho_{12}^{2}}{\rho_{12}^{2}-u^{2}} \frac{1-\rho_{34}^{2}}{\rho_{34}^{2}-u^{2}} } \right) \, .
\end{aligned}
\end{equation}
The above provides an exact result for the parallel correlation function when order parameter fields are placed along the interface, as shown in Fig.~\ref{fig_ssss_a}.

In the following, we further specialize the parallel correlation function to spin fields symmetrically arranged with respect to the horizontal axis. We set $y_{1}=-y_{4}$, $y_{2}=-y_{3}$ and observe that this choice implies $\rho_{12}=\rho_{14}$ in (\ref{12042021_0843}) and (\ref{a011a}). Thus, we introduce
\begin{equation}
\label{ }
G_{\parallel}^{(\rm sym)}(y_{1},y_{2}) = G_{\parallel}(y_{1},y_{2},-y_{2},-y_{1}) \, ,
\end{equation}
the above reads
\begin{equation}
\begin{aligned}
\label{31122020_0905}
G_{\parallel}^{(\rm sym)}(y_{1},y_{2}) & = \widetilde{ \langle \sigma_{i} \rangle } ^{4} + \frac{1}{4} \widetilde{ \langle \sigma_{i} \rangle }^{2} \left( \Delta \langle \sigma_{i} \rangle \right)^{2} \mathcal{C}_{\parallel}^{(\rm sym)}(\tau_{1},\tau_{2}) + \frac{1}{16} \left( \Delta \langle \sigma_{i} \rangle \right)^{4} \mathcal{G}_{\parallel}^{(\rm sym)}(\tau_{1},\tau_{2}) + \Os(R^{-1/2}) \, ,
\end{aligned}
\end{equation}
with the scaling functions
\begin{equation}
\begin{aligned}
\label{31122020_0906}
\mathcal{C}_{\parallel}^{(\rm sym)}(\tau_{1},\tau_{2}) & = \frac{2}{\pi} \biggl[ 2 \sin^{-1}(\rho_{12}) + 2 \sin^{-1}(\rho_{12}\rho_{23}) + \sin^{-1}(\rho_{23}) + \sin^{-1}(\rho_{12}^{2}\rho_{23}) \biggr] \, ,
\end{aligned}
\end{equation}
and
\begin{equation}
\begin{aligned}
\label{31122020_0907}
\mathcal{G}_{\parallel}^{(\rm sym)}(\tau_{1},\tau_{2}) & = \frac{4}{\pi^{2}} \left( \sin^{-1}(\rho_{12}) \right)^{2} + \frac{4}{\pi^{2}} \int_{0}^{\rho_{12}^{2}\rho_{23}} \frac{\textrm{d}u}{\sqrt{1-u^{2}}} \sin^{-1}\left( u \frac{1-\rho_{12}^{2}}{\rho_{12}^{2}-u^{2}} \right) \, .
\end{aligned}
\end{equation}

We defer the examination of some limiting cases in which the parallel correlation function reduces to the rhomboidal or rectangular one in Sec.~\ref{section32} and Sec.~\ref{section33}, respectively.

%=====================================================================================
\subsection{Rhomboidal correlation function}
\label{section32}
The analysis for the correlation function with spin fields in rhomboidal arrangement does not proceed along the lines illustrated in the previous section. Firstly, because we don't provide a general expression for $F_{4}(X_{1},X_{2},X_{3},X_{4} \vert \textsf{R}_{1234})$ which we could use in the limit of interest, secondly because the correlation coefficient degenerates to unity, i.e., $\rho_{23} =1$; such a limit has to be handled carefully. The perfect correlation between the random variables $u_{2}$ and $u_{3}$ -- as signaled by $\rho_{23} =1$ -- is best treated by taking the limit $\rho_{23} \rightarrow 1$ in the passage probability density (\ref{11032021_1534}). The correlation function then follows by plugging the passage probability for the rhomboidal arrangement, hereinafter denoted $P_{4}^{(\diamond)}$, into (\ref{a006a}). This is the calculation strategy which we are going to detail.

The passage probability for the rhomboidal arrangement is obtained from the following limiting procedure
\begin{equation}
\label{30122020_1055}
P_{4}^{(\diamond)} = \lim_{\rho_{23}\rightarrow1} P_{4} \, .
\end{equation}
Since $y \equiv y_{1}=-y_{4}$, we have $\rho_{12}=\rho_{34} \equiv \alpha$, with
\begin{equation}
\label{12042021_0913}
\alpha=\sqrt{\frac{1-\tau}{1+\tau}} \, , \qquad \tau = \frac{2y}{R} \, .
\end{equation}
As a result, the passage probability $P_{4}^{(\diamond)}$ is characterized by one correlation coefficient, say $\alpha$; note that $\rho_{14}=\rho_{12}\rho_{23}\rho_{34}=\alpha^{2}$ by virtue of $\rho_{23}=1$. Recalling the relationship between $P_{4}$ and $\Pi_{4}$ given by (\ref{11032021_1534}), a simple calculation yields for the rescaled passage probability the following expression
\begin{equation}
\label{30122020_1056}
\Pi_{4}^{(\diamond)}(u_{1},u_{2},u_{3},u_{4} \vert \alpha) = \frac{\sqrt{2\pi}}{4\pi^{2}(1-\alpha^{2})} \delta(u_{2}-u_{3}) \textrm{e}^{-\frac{Q}{2(1-\alpha^{2})}} \, ,
\end{equation}
where
\begin{equation}
\label{ }
Q = u_{1}^{2} + u_{4}^{2} + \alpha^{2} (u_{2}^{2}+u_{3}^{2}) + (1-\alpha^{2}) u_{2} u_{3} - 2\alpha (u_{1}u_{2}+u_{3}u_{4}) \, .
\end{equation}
A simple check reveals that $\Pi_{4}^{(\diamond)}(u_{1},u_{2},u_{3},u_{4} \vert \alpha)$ is correctly normalized. This fact can be confirmed by a direct calculation:
\begin{equation}
\label{ }
\int_{\mathbb{R}^{2}} \textrm{d}u_{1} \textrm{d}u_{4} \, \Pi_{4}^{(\diamond)}(u_{1},u_{2},u_{3},u_{4} \vert \alpha) = \frac{1}{\sqrt{2\pi}} \delta(u_{2}-u_{3}) \textrm{e}^{-u_{2}u_{3}/2} \, ,
\end{equation}
thus
\begin{equation}
\label{ }
\int_{\mathbb{R}^{3}} \textrm{d}u_{1} \textrm{d}u_{3} \textrm{d}u_{4} \, \Pi_{4}^{(\diamond)}(u_{1},u_{2},u_{3},u_{4} \vert \alpha) = \frac{1}{\sqrt{2\pi}} \textrm{e}^{-u_{2}^{2}/2} = \Pi_{1}(u_{2}) \, ,
\end{equation}
and since $\Pi_{1}(u_{2})$ is normalized, it follows that
\begin{equation}
\label{ }
\int_{\mathbb{R}^{4}} \textrm{d}u_{1} \textrm{d}u_{2} \textrm{d}u_{3} \textrm{d}u_{4} \, \Pi_{4}^{(\diamond)}(u_{1},u_{2},u_{3},u_{4} \vert \alpha) = 1 \, ,
\end{equation}
as we anticipated and as consistency requires.

The four-point correlation function is computed by plugging the passage probability (\ref{30122020_1056}) into the probabilistic representation (\ref{a006a}). Leaving in Appendix \ref{Appendix_A} the lengthy mathematical manipulations, the result for the rhomboidal correlation function reads
\begin{equation}
\begin{aligned}
\label{31122020_0843}
G_{\diamond}(x,y) & = \widetilde{\langle\sigma_{i}\rangle}^{4} + \frac{1}{4} \left( \Delta\langle\sigma_{i}\rangle \right)^{2} \widetilde{\langle\sigma_{i}\rangle}^{2} \mathcal{C}_{\diamond}(\eta,\tau) + \frac{1}{16} \left( \Delta\langle\sigma_{i}\rangle \right)^{4} \mathcal{G}_{\diamond}(\eta,\tau) + \Os(R^{-1/2}) \, ,
\end{aligned}
\end{equation}
where $\mathcal{C}_{\diamond}(\eta,\tau)$ and $\mathcal{G}_{\diamond}(\eta,\tau)$ are the scaling functions
\begin{equation}
\begin{aligned}
\label{11032021_1716}
\mathcal{C}_{\diamond}(\eta,\tau) & = 1 - 2 \textrm{erf}(|\eta|) + \frac{2}{\pi} \sin^{-1}\left( \alpha^{2} \right) + 16 T(\sqrt{2}\eta,r) \\
\mathcal{G}_{\diamond}(\eta,\tau) & = \frac{2}{\pi} \sin^{-1}\left( \alpha^{2} \right) - 2 \mathcal{Y}(|\eta|,r) \, ,
\end{aligned}
\end{equation}
where $T$ is Owen's function (see (\ref{31122020_0850}) for its definition) and $\mathcal{Y}(\eta,r)$ is defined by
\begin{equation}
\label{31122020_0849}
\mathcal{Y}(\eta,r) = \frac{2}{\sqrt{\pi}} \int_{0}^{\eta}\textrm{d}u \, \textrm{e}^{-u^{2}} \textrm{erf}^{2}(ru) \, ,
\end{equation}
where $\textrm{erf}(z)=(2/\sqrt{\pi})\int_{0}^{z}\textrm{d}t \, \textrm{e}^{-t^{2}}$ is the error function \cite{Temme}, $\eta=x/\lambda$ is the rescaled horizontal coordinate, and
\begin{equation}
\label{12042021_0914}
r = \frac{\alpha}{\sqrt{1-\alpha^{2}}} \, .
\end{equation}

Let us discuss some general properties of the rhomboidal correlation function. By definition, $G_{\diamond}(x,y)$ is an even function under the exchange of $x$ with $-x$. The parity $G_{\diamond}(x,y)=G_{\diamond}(-x,y)$ is thus satisfied by both the scaling functions (\ref{11032021_1716}); recall that $T(\sqrt{2}\eta,r)=T(-\sqrt{2}\eta,r)$ (see (\ref{31122020_0850})). The limit $x\rightarrow+\infty$ for the rhomboidal correlation function projects the order parameter field $\sigma_{i}(x, 0)$ deep into the $b$ phase and, analogously, $\sigma_{i}(-x, 0)$ deep into the $a$ phase. As a result, we obtain the following clustering relation
\begin{equation}
\begin{aligned}
\label{09042021_1542}
\lim_{x \rightarrow + \infty} G_{\diamond}(x,y) & = \langle \sigma_{i} \rangle_{a} \langle \sigma_{i} \rangle_{b} \biggl[ \left( \frac{ \langle \sigma_{i} \rangle_{a} + \langle \sigma_{i} \rangle_{b} }{ 2 } \right)^{2} + \left( \frac{ \langle \sigma_{i} \rangle_{a} - \langle \sigma_{i} \rangle_{b} }{ 2 } \right)^{2} \frac{2}{\pi} \sin^{-1}\left( \alpha^{2} \right) \biggr] 
 \, .
\end{aligned}
\end{equation}
The quantity enclosed in square brackets is the two-point correlation function with order parameter fields placed along the interface, i.e., $\langle \sigma_{i}(0,y) \sigma_{i}(0,-y) \rangle_{ab}$ \cite{DS_twopoint}.

As a cautionary check, it is useful to compare the limit $x \rightarrow 0$ of the rhomboidal correlation function with the limit $y_{2} \rightarrow 0$ of the parallel correlation function. Consistency requires that the above limits have to coincide, i.e.
\begin{equation}
\label{31122020_0853}
\lim_{x \rightarrow 0} G_{\diamond}(x,y) = \lim_{y_{2} \rightarrow 0} G_{\parallel}^{(\rm sym)}(y,y_{2}) \, .
\end{equation}
In order to check that (\ref{31122020_0853}) is satisfied, we firstly examine the limit $x \rightarrow 0$ of the scaling functions which appear in the correlation function $G_{\diamond}(x,y)$. For the scaling function $\mathcal{C}_{\diamond}(\eta, \tau)$, we have
\begin{equation}
\begin{aligned}
\label{12032021_1125}
\lim_{x \rightarrow 0 } \mathcal{C}_{\diamond}(\eta, \tau) & = 1 + \frac{2}{\pi} \sin^{-1}\left( \alpha^{2} \right) + \frac{8}{\pi}\tan^{-1}\left( \frac{\alpha}{\sqrt{1-\alpha^{2}}} \right) \\
& = 1 + \frac{2}{\pi} \sin^{-1}\left( \alpha^{2} \right) + \frac{8}{\pi}\sin^{-1}(\alpha) \, ;
\end{aligned}
\end{equation}
in the first line, we used property (\ref{31122020_0855}) of Owen's $T$-function while in the second line, we used the trigonometric identity $\tan^{-1}\left( \alpha/\sqrt{1-\alpha^{2}} \right)=\sin^{-1}(\alpha)$. The scaling function $\mathcal{G}_{\diamond}(\eta, \tau)$ reduces to
\begin{equation}
\begin{aligned}
\label{24042021_1420}
\lim_{x \rightarrow 0 } \mathcal{G}_{\diamond}(\eta, \tau) & = \frac{2}{\pi} \sin^{-1}\left( \alpha^{2} \right) \, .
\end{aligned}
\end{equation}
Let us consider now the right hand side of (\ref{31122020_0853}). The limit $y_{2} \rightarrow 0$ of the scaling function $\mathcal{C}_{\parallel}^{(\rm sym)}(\tau,\tau_{2})$ gives (see (\ref{31122020_0906}))
\begin{equation}
\begin{aligned}
\label{24042021_1409}
\lim_{y_{2} \rightarrow 0} \mathcal{C}_{\parallel}^{(\rm sym)}(\tau, \tau_{2}) & = \lim_{\rho_{12} \rightarrow \alpha}  \lim_{\rho_{23} \rightarrow 1} \mathcal{C}_{\parallel}^{(\rm sym)}(\tau, \tau_{2}) \\
& = 1 + \frac{2}{\pi} \sin^{-1}\left( \alpha^{2} \right) + \frac{8}{\pi}\sin^{-1}(\alpha) \, ,
\end{aligned}
\end{equation}
which coincides (\ref{12032021_1125}), as it should. Finally, we consider the same limit discussed above but now for scaling function $\mathcal{G}_{\parallel}^{(\rm sym)}(\tau,\tau_{2})$. From (\ref{31122020_0907}), we have
\begin{equation}
\begin{aligned}
\label{24042021_1426}
\lim_{y_{2} \rightarrow 0} \mathcal{G}_{\parallel}^{(\rm sym)}(\tau, \tau_{2}) & = \lim_{\rho_{12} \rightarrow \alpha}  \lim_{\rho_{23} \rightarrow 1} \mathcal{G}_{\parallel}^{(\rm sym)}(\tau, \tau_{2}) \\
& = \frac{4}{\pi^{2}} \left( \sin^{-1}(\alpha) \right)^{2} + \frac{4}{\pi^{2}} \int_{0}^{\alpha^{2}} \frac{\textrm{d}u}{\sqrt{1-u^{2}}} \sin^{-1}\left( u \frac{1-\alpha^{2}}{\alpha^{2}-u^{2}} \right) \\
& = \frac{2}{\pi} \sin^{-1}\left( \alpha^{2} \right) \, ,
\end{aligned}
\end{equation}
the last equality follows from the identity (\ref{idA13}), whose proof is supplied in Appendix \ref{Appendix_A}. Summarizing, thanks to the results (\ref{12032021_1125})-(\ref{24042021_1426}), we have proved the connection between the parallel and rhomboidal correlation functions given by (\ref{31122020_0853}).

\begin{figure}[htbp]
\centering
\includegraphics[width=105mm]{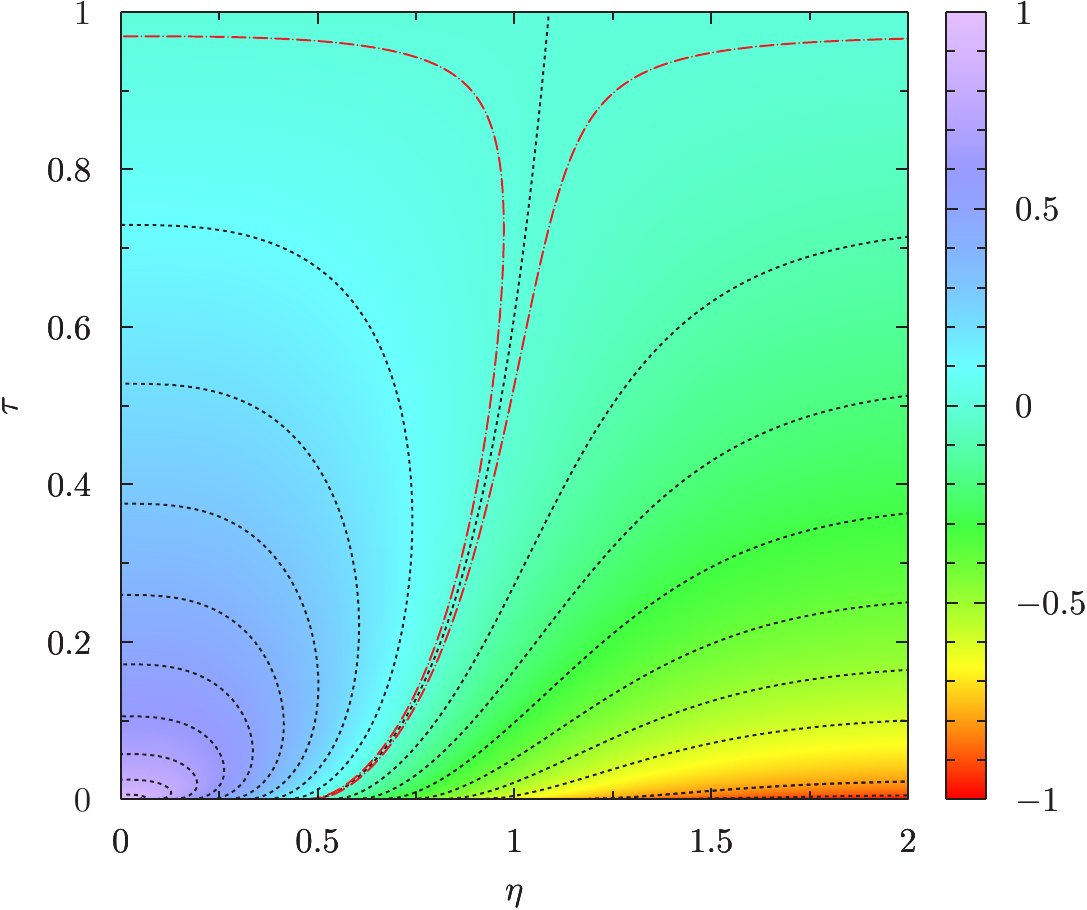}
\caption{The scaling function $\mathcal{G}_{\diamond}(\eta,\tau)$. The dot-dashed red curves indicate the contour lines corresponding to the values $\pm 0.01$, the black dashed curve between the red ones is the contour line where $\mathcal{G}_{\diamond}(\eta,\tau)$ vanishes. The remaining black dashed curves are contour lines corresponding to the values ranging from $-0.9$ to $+0.9$ with spacing $0.1$.}
\label{fig_rhombusdensity}
\end{figure}

The scaling function $\mathcal{G}_{\diamond}(\eta,\tau)$ is plotted in Fig.~\ref{fig_rhombusdensity} as function of $\eta$ and $\tau$.

%=====================================================================================
\subsection{Rectangular correlation function}
\label{section33}
The correlation function with spin fields arranged as shown in Fig.~\ref{fig_ssss_c} can be calculated by following the procedure outlined in Sec.~\ref{section32}. Leaving the mathematical details in Appendix \ref{Appendix_A}, the rectangular correlation function admits the following representation
\begin{equation}
\begin{aligned}
\label{11032021_1305}
G_{\oblong}(x,y) & =  \widetilde{\langle\sigma_{i}\rangle}^{4} + \frac{1}{4} \left( \Delta\langle\sigma_{i}\rangle \right)^{2} \widetilde{\langle\sigma_{i}\rangle}^{2} \mathcal{C}_{\oblong}(\eta,\tau) + \frac{1}{16} \left( \Delta\langle\sigma_{i}\rangle \right)^{4} \mathcal{G}_{\oblong}(\eta,\tau) + \Os(R^{-1/2}) \, ,
\end{aligned}
\end{equation}
with the scaling functions
\begin{equation}
\begin{aligned}
\label{11032021_1717}
\mathcal{C}_{\oblong}(\eta,\tau) & = 2 - 4 \textrm{erf}(|\chi|) - 16 T(\sqrt{2}\chi,\sqrt{\tau}) + 16 T(\sqrt{2}\chi,1/\sqrt{\tau}) \\
\mathcal{G}_{\oblong}(\eta,\tau)  & =  5 - 4 \textrm{erf}(|\chi|) - 16 T(\sqrt{2}\chi,\sqrt{\tau}) - 16 T(\sqrt{2}\chi,1/\sqrt{\tau}) \, ,
\end{aligned}
\end{equation}
and the rescaled coordinates $\eta=x/\lambda$, $\chi=\eta/\kappa$, $\kappa=\sqrt{1-\tau^{2}}$, $\tau=2y/R$.

Analogously to the rhomboidal case, reflection symmetry is expected, namely: $G_{\oblong}(x,y)=G_{\oblong}(-x,y)$. The limit $x \rightarrow + \infty$ in (\ref{11032021_1305}) entails a separation of the four spin fields into two infinitely separated clusters. By using the properties of Owen's $T$ function listed below (\ref{31122020_0850}) and the limits
\begin{eqnarray}
\label{}
\lim_{x \rightarrow +\infty } \mathcal{C}_{\oblong}(\eta, \tau) & = & - 2  \\
\lim_{x \rightarrow +\infty } \mathcal{G}_{\oblong}(\eta, \tau) & = & 1 \, ,
\end{eqnarray}
we have
\begin{equation}
\begin{aligned}
\label{09042021_1503}
\lim_{x \rightarrow + \infty} G_{\oblong}(x,y) & = \langle \sigma_{i} \rangle_{a}^{2} \langle \sigma_{i} \rangle_{b}^{2} \, ,
\end{aligned}
\end{equation}
which is the expected clustering behavior. Then, the limit $x \rightarrow 0$ of the rectangular correlation function must reduce to the limit $y_{2}\rightarrow y_{1} \equiv y$ in the parallel correlation function, i.e.,
\begin{equation}
\begin{aligned}
\label{24042021_1341}
\lim_{x \rightarrow 0} G_{\oblong}(x,y) & = \lim_{y_{2} \rightarrow y} G_{\parallel}^{(\rm sym)}(y,y_{2}) \, ,
\end{aligned}
\end{equation}
which is indeed the case. The identity (\ref{24042021_1341}) can be easily verified by observing that the scaling functions which appear in the left hand side of (\ref{24042021_1341}) satisfy the limits
\begin{eqnarray}
\label{24042021_1350}
\lim_{x \rightarrow 0 } \mathcal{C}_{\oblong}(\eta, \tau) & = & 6 - \frac{16}{\pi} \tan^{-1}(\sqrt{\tau}) \\
\lim_{x \rightarrow 0 } \mathcal{G}_{\oblong}(\eta, \tau) & = & 1 \, ,
\end{eqnarray}
while for the scaling functions in the right hand side, we have
\begin{eqnarray}
\label{24042021_1351}
\lim_{y_{2} \rightarrow y } \mathcal{C}_{\parallel}^{(\rm sym)}(\tau, \tau_{2}) & = & 2 + \frac{8}{\pi} \sin^{-1}\left( \frac{1-\tau}{1+\tau} \right) \\
\lim_{y_{2} \rightarrow y } \mathcal{G}_{\parallel}^{(\rm sym)}(\tau, \tau_{2}) & = & 1 \, .
\end{eqnarray}
By virtue of the identity
\begin{equation}
\label{ }
\frac{2}{\pi} \sin^{-1}\left( \frac{1-\tau}{1+\tau} \right) = 1 - \frac{4}{\pi} \tan^{-1}(\sqrt{\tau}) \, ,
\end{equation}
the right hand side of the first equations in (\ref{24042021_1350}) and (\ref{24042021_1351}) coincide, and the property (\ref{24042021_1341}) follows.

\begin{figure}[htbp]
\centering
\includegraphics[width=105mm]{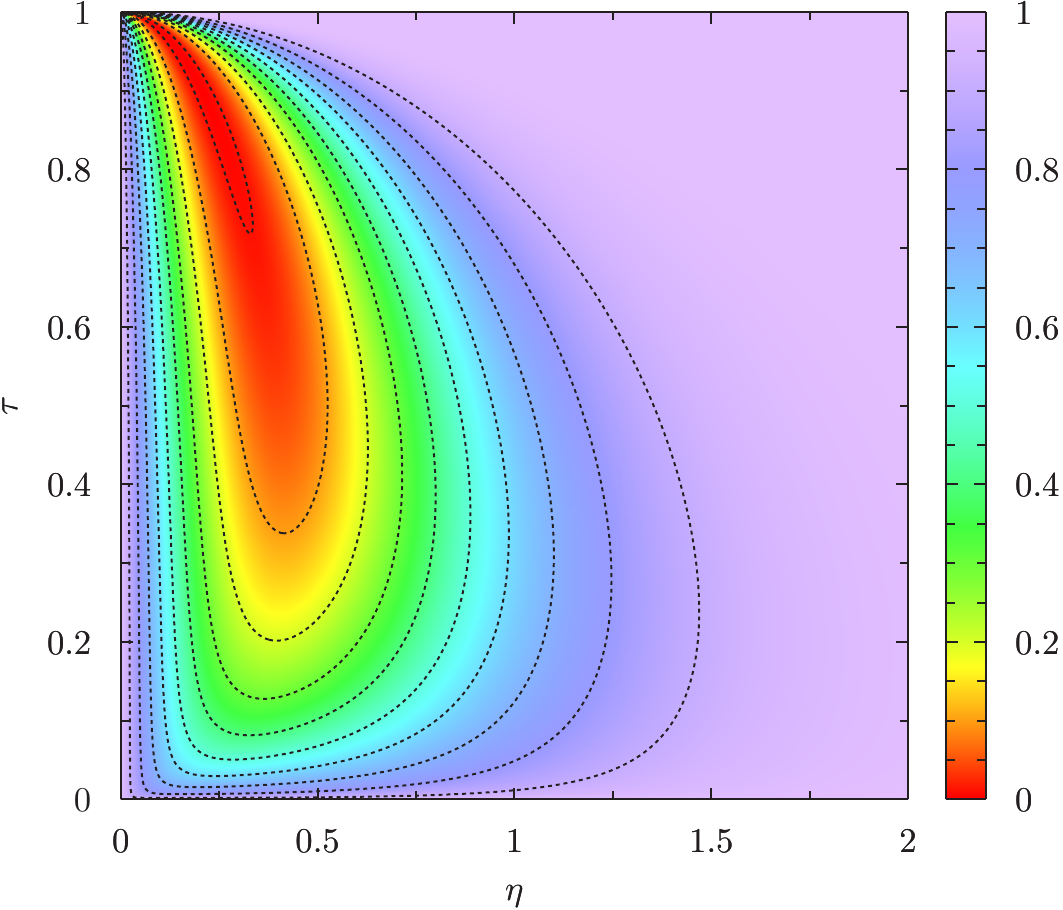}
\caption{The scaling function $\mathcal{G}_{\oblong}(\eta,\tau)$. Dashed curves indicate contour lines corresponding to the values ranging from $0.9$ to $0.1$ with spacing $0.1$ from outer to next to inner curve, and $0.01$ for the inner curve.}
\label{fig_rectangledensity}
\end{figure}

The analytical examination of interfacial correlations can be carried out by taking the limit of small $\tau$ in (\ref{11032021_1717}). Among the two scaling functions in (\ref{11032021_1717}), we focus on $\mathcal{G}_{\oblong}(\eta,\tau)$ because it is the one which characterizes the Ising model. The analysis we are going to provide for $\mathcal{G}_{\oblong}(\eta,\tau)$ can be straightforwardly extended to the scaling function $\mathcal{C}_{\oblong}(\eta,\tau)$. The analytical study is better achieved by writing the scaling function $\mathcal{G}_{\oblong}(\eta,\tau)$ in the equivalent form
\begin{equation}
\label{09042021_1815}
\mathcal{G}_{\oblong}(\eta,\tau) =  1 - 4 \textrm{erf}(\chi)\textrm{erfc}(\chi/\sqrt{\tau}) - 16 T(\sqrt{2}\chi,\sqrt{\tau}) + 16 T(\sqrt{2}\chi/\sqrt{\tau},\sqrt{\tau}) \, ,
\end{equation}
which follows from the functional identity (\ref{09042021_1657}) satisfied by the $T$-function; in the above, $\textrm{erfc}(z)=1-\textrm{erf}(z)$ is the complementary error function \cite{Temme}. The scaling function $\mathcal{G}_{\oblong}(\eta,\tau)$ is plotted in Fig.~\ref{fig_rectangledensity}. As already anticipated by the study of clustering properties, $\mathcal{G}_{\oblong}(\eta,\tau)$ exhibits a nontrivial dependence through the coordinates $\eta$ and $\tau$ only within the interfacial region, i.e., when the rescaled horizontal coordinate $\eta$ is not large and the vertical one does not vanish ($\tau>0$).

For small vertical separations $\tau$ the scaling function (\ref{09042021_1815}) can be approximated with the expansion
\begin{equation}
\label{09042021_1944}
\mathcal{G}_{\oblong}(\eta,\tau) = 1 - 4 \textrm{erf}(\eta)\textrm{erfc}(\eta/\sqrt{\tau}) + \frac{16}{2\pi} \sqrt{\tau} \Bigl[ \textrm{e}^{-\eta^{2}/\tau} - \textrm{e}^{-\eta^{2}} \Bigr] + \Os(\tau^{3/2}) \, ,
\end{equation}
up to higher-order corrections of order $\Os(\tau^{3/2})$. The small-$y$ expansion provided by (\ref{09042021_1944}) accurately encompasses any value of $\eta$ both within and without the interfacial region provided $\xi \ll y \ll R$. The term $\sqrt{\tau} \propto \sqrt{y/R}$ is the signature of long range correlations in the direction parallel to the interface. The term enclosed in square brackets provides a non-trivial spatial dependence which is specific for the rectangular arrangement of spin fields.

The $T$-functions in (\ref{09042021_1815}) may not allow for a direct visualization of the dependence through the horizontal coordinate $\eta$, the coordinate transverse to the interface. Alternatively, we can examine the slope in the horizontal direction, the latter is given by the first derivative of $\mathcal{G}_{\oblong}(\eta,\tau)$ with respect to $\eta$
\begin{equation}
\label{18042021_0457}
\partial_{\eta} \mathcal{G}_{\oblong}(\eta,\tau) = \frac{8}{\sqrt{\pi}\kappa} \textrm{e}^{-\chi^{2}} \bigl[ \textrm{erf}(\sqrt{\tau}\chi) - \textrm{erfc}(\chi/\sqrt{\tau}) \bigr] \, .
\end{equation}
The overall exponential envelope $\textrm{e}^{-\chi^{2}}$ suppresses interfacial correlations far away from the interfacial region ($|\chi| \gg1$) and determines the confinement of the long-ranged character of density fluctuations in the region where $|\chi| \ll 1$, where it is more probable to find the interface.

We conclude this section with a discussion on a general property satisfied by correlation functions. It is known from rigorous studies that fluctuations of the interface midpoint grow as $\sqrt{R}$ \cite{Gallavotti_72, Gallavotti} and that interface fluctuations in Ising \cite{GI} and  $q$-state Potts models \cite{CIV} are characterized by the Brownian bridge property. This feature is also well visualized by the explicit form of the passage probability $P_{1}(x,0)=\pi^{-1/2}\lambda^{-1}\textrm{e}^{-(x/\lambda)^{2}}$, which gives for the mean square position of the interface midpoint
\begin{equation}
\label{ }
\mathbb{E}[x^{2}] = \int_{\mathbb{R}}\textrm{d}x \, x^{2} P_{1}(x,0) = \frac{R\xi}{2} \, ,
\end{equation}
meaning that $\langle x \rangle \equiv \sqrt{\mathbb{E}[x^{2}]} \propto \sqrt{R}$. In the limit $R \rightarrow \infty$ with fixed $x$ and $y$ both the rescaled variables variables $\eta=x/\sqrt{R\xi}$ and $\tau=2y/R$ tend to zero. As a result, the scaling functions encountered in this paper satisfy the following relations
\begin{equation}
\label{ }
\lim_{R \rightarrow \infty} \mathcal{C}_{q} = 6 \, , \qquad \lim_{R \rightarrow \infty} \mathcal{G}_{q} = 1 \, , \qquad q \in \{\parallel, \diamond, \oblong \} \, ,
\end{equation}
which can be obtained by evaluating the scaling functions for $\eta=\tau=0$. The above limits imply
\begin{equation}
\begin{aligned}
\label{}
\lim_{R \rightarrow + \infty} \langle \sigma_{i}(x_{1},y_{1}) \sigma_{i}(x_{2},y_{2}) \sigma_{i}(x_{3},y_{3}) \sigma_{i}(x_{4},y_{4}) \rangle_{ab} & = \widetilde{\langle\sigma_{i}\rangle}^{4} + \frac{6}{4} \widetilde{\langle\sigma_{i}\rangle}^{2} \left( \Delta\langle\sigma_{i}\rangle \right)^{2} + \frac{1}{16} \left( \Delta\langle\sigma_{i}\rangle \right)^{4} \\
& = \frac{ \langle \sigma_{i} \rangle_{a}^{4} + \langle \sigma_{i} \rangle_{b}^{4} }{ 2 } \, ,
\end{aligned}
\end{equation}
meaning that the limit $R \rightarrow \infty$ of the four point correlation function yields an averaging over coexisting phases. The above limit is actually a particular case of a more general result which holds for any $n$ and generic universality classes \cite{Squarcini_Multipoint}, and includes an analogous result for the Ising model \cite{Abraham_review} as a particular case.

%=====================================================================================
\section{Numerical results}
\label{sec3}
The analytical predictions obtained in the previous sections are specialized to the Ising model and compared against Monte Carlo simulations. We set the notation by recalling the lattice Hamiltonian
\begin{equation}
\label{ }
\mathcal{H} = - J \sum_{ \langle i,j \rangle} s_{i} s_{j} \, ,
\end{equation}
where $s_{i} = \pm 1$ and the sum is restricted over nearest neighboring sites of a two-dimensional square lattice. The critical temperature is given by $T_{\textrm{c}}/J=2/\log(1+\sqrt{2})=2.269\,185\dots$ \cite{Onsager_44}. The correlation length\footnote{Lengths are measured in units of lattice spacing.} in the low-temperature phase reads
\begin{equation}
\label{2022}
\xi = (4K-4K^{\star})^{-1} \, ,
\end{equation}
with the dual coupling $K^{\star}$ defined by means of $\exp(-2K^{\star})=\tanh K$ with $K=J/T$ \cite{Abraham_review}. Without loss of generality, we set the ferromagnetic coupling $J=1$ in our simulations. 

We consider boundary conditions with $a=-1$, $b=+1$ and recall that Ising symmetry implies $\langle \sigma \rangle_{+} = - \langle \sigma \rangle_{-}$. The spontaneous magnetization $M=\langle \sigma \rangle_{+}>0$ is given by \cite{McCoyWu,Yang}
\begin{equation}
\label{2032}
M = \left( 1-(\sinh(2K))^{-4} \right)^{1/8} \, .
\end{equation}
From the above boundary conditions it follows that $\widetilde{\langle \sigma \rangle}=0$ and $\Delta \langle \sigma \rangle=-2M$. Focusing on the leading-order result, the parallel, rhomboidal and rectangular correlation functions reduce to:
\begin{equation}
\begin{aligned}
\label{11032021_1715}
G_{\parallel}^{(\rm sym)}(y_{1},y_{2}) & = M^{4} \mathcal{G}_{\parallel}^{(\rm sym)}(\tau_{1},\tau_{2}) \\
G_{\diamond}(x,y) & = M^{4} \mathcal{G}_{\diamond}(\eta,\tau) \\
G_{\oblong}(x,y) & = M^{4} \mathcal{G}_{\oblong}(\eta,\tau) \, ,
\end{aligned}
\end{equation}
with $\mathcal{G}_{\parallel}$, $\mathcal{G}_{\diamond}$, and $\mathcal{G}_{\oblong}$ the scaling functions  (\ref{31122020_0907}), (\ref{11032021_1716}), and (\ref{11032021_1717}), respectively. We recall that $\rho_{12}$ and $\rho_{23}$ in (\ref{31122020_0907}) are given by (\ref{04042021_2028}) with $\tau_{2}=-\tau_{3}$. Then, $\eta$ and $\tau$ in (\ref{11032021_1715}) are defined below (\ref{11032021_1717}).

We emphasize that (\ref{11032021_1715}) are exact results at leading order in finite-size corrections. Interface structure corrections proportional to $R^{-1/2}$ occur in general but actually vanish for the Ising model \cite{Squarcini_Multipoint}. As a result, (\ref{11032021_1715}) are valid up to subleading corrections at order $\Os(R^{-1})$. 

The numerical simulations have been carried out on a rectangular lattice of horizontal length $L$ and temperature $T$ such that $\xi \ll R$. The theory is defined for $L\rightarrow \infty$ but in practice it is enough to take $L/\lambda \gtrsim 7$ as a criterion for the sizing of the simulation box. We summarize the details of the simulation scheme which we have already employed in a companion paper \cite{ST_threepoint}. The simulations are performed by means of a hybrid Monte Carlo scheme (see, e.g. \cite{LandauBinder}) which combines the standard Metropolis algorithm and the Wolff cluster algorithm \cite{Wolff}. The minimum number of MC steps per site is $10^{7}$. Parallelization was obtained by independently and simultaneously simulating up to 128 Ising lattices on a parallel computer. An appropriately seeded family of dedicated, very large period, Mersenne Twister random number generators \cite{MN_1998b}, in the MT2203 implementation of the Intel Math Kernel Library, was used in order to simultaneously generate independent sequences of random number to be used for the MC updates of the lattices.

\begin{figure}[htbp]
\centering
\includegraphics[width=105mm]{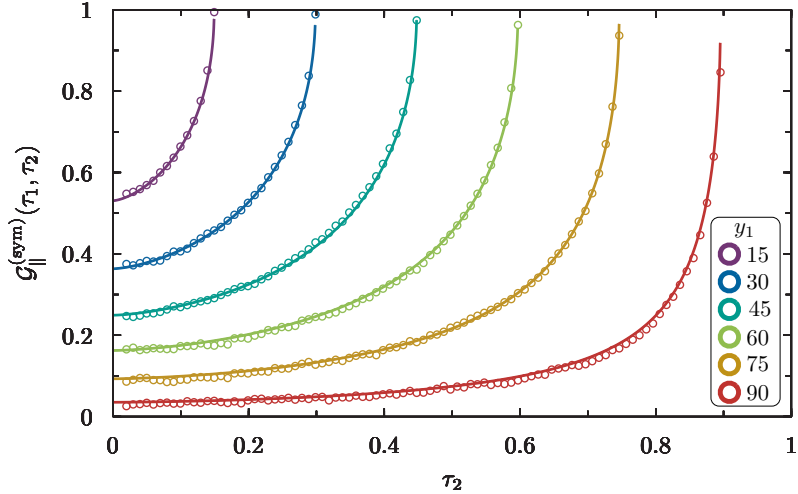}
\caption{Parallel correlation function $\mathcal{G}_{\parallel}^{(\rm sym)}(\tau_{1},\tau_{2})$ as function of $\tau_{2}$ for fixed $\tau_{1}=2y_{1}/R$. Data points are obtained from numerical simulations at $T=2$, $R=201$.}
\label{fig_MC1}
\end{figure}

The first observable we consider is the parallel correlation function. The analytic result (\ref{31122020_0907}) and the numerical data are compared in Fig.~\ref{fig_MC1} as function of $\tau_{2}=2y_{2}/R$ for fixed $\tau_{1}=2y_{1}/R$. For small values of $y_{2}$ the correlation function approaches the asymptotic value given by
\begin{equation}
\begin{aligned}
\label{}
\lim_{y_{2} \rightarrow 0 } \mathcal{G}_{\parallel}^{(\rm sym)}(\tau_{1},\tau_{2}) & = \frac{2}{\pi}\sin^{-1}\left( \frac{1-\tau_{1}}{1+\tau_{1}} \right) \, .
\end{aligned}
\end{equation}
The above result is actually confirmed by the numerical simulations. For each value of $y_{1}$ sampled in Fig.~\ref{fig_MC1}, $\tau_{1}$ provides the maximum value allowed for $\tau_{2}$. This maximum value is reached when the order parameter field $\sigma(0,y_{2})$ approaches $\sigma(0,y_{1})$ from below; see Fig.~\ref{fig_ssss_a}. In such a regime the correlation function approaches the limiting value
\begin{equation}
\label{ }
\lim_{y_{2} \rightarrow y_{1} } \mathcal{G}_{\parallel}^{(\rm sym)}(\tau_{1},\tau_{2}) = 1 \, ,
\end{equation}
a feature which is well visible in the plot of Fig.~\ref{fig_MC1}. It has to be emphasized that such a limiting result is obtained in field theory from the single-particle contribution which dominates the asymptotic of the large-separation behavior of correlation functions and that further corrections arising from interface structure effects are expected at order $\Os(R^{-1})$ for the Ising model \cite{Squarcini_Multipoint}. Therefore the limit $y_{2} \rightarrow y_{1}$ has to be understood in the sense that the separation $y_{1}-y_{2}$ is small compared to $R$ but it has to be large compared to the bulk correlation length, i.e., $\xi \ll y_{1}-y_{2} \ll R$. With this in mind, we actually observe an excellent agreement between theory and numerics.

\begin{figure*}[htbp]
\centering
%\hspace{30mm}
        \begin{subfigure}[b]{0.49\textwidth}
            \centering
            \includegraphics[height=47mm]{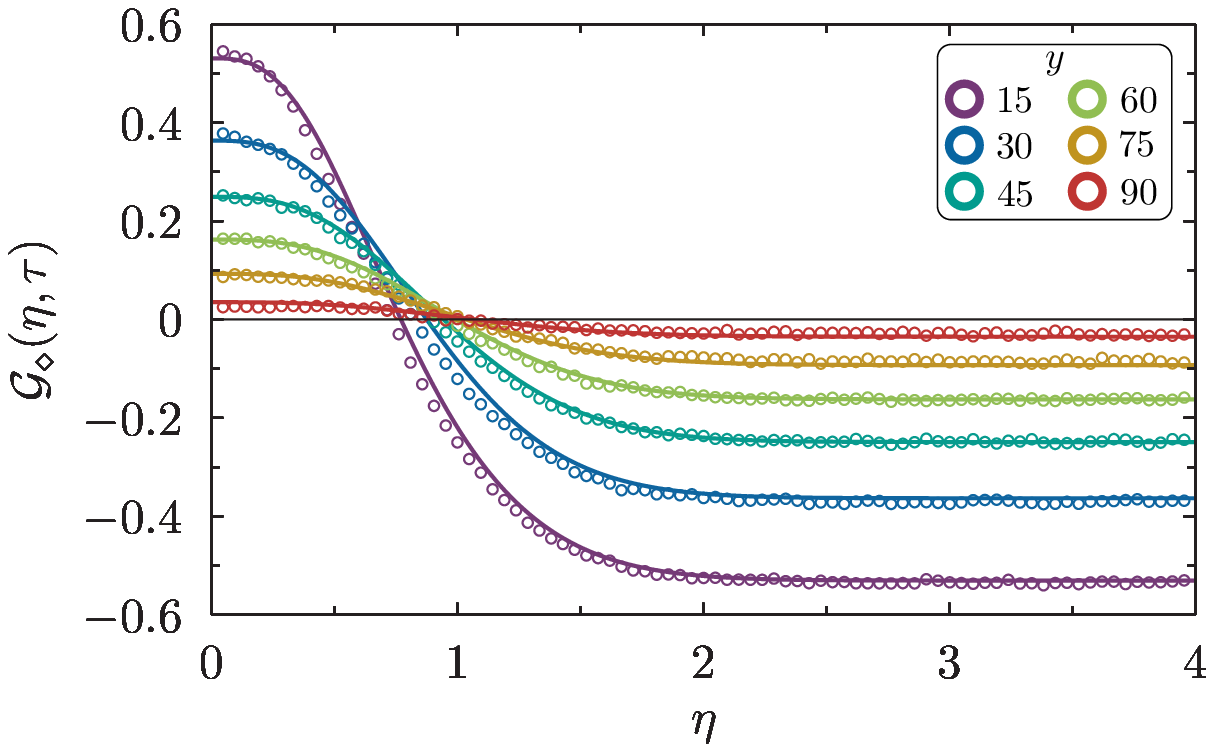}
            \caption[]%
            {{\small }}    
            \label{fig_rh_eta}
        \end{subfigure}
\hfill
        \begin{subfigure}[b]{0.50\textwidth}  
            \centering 
            \includegraphics[height=47mm]{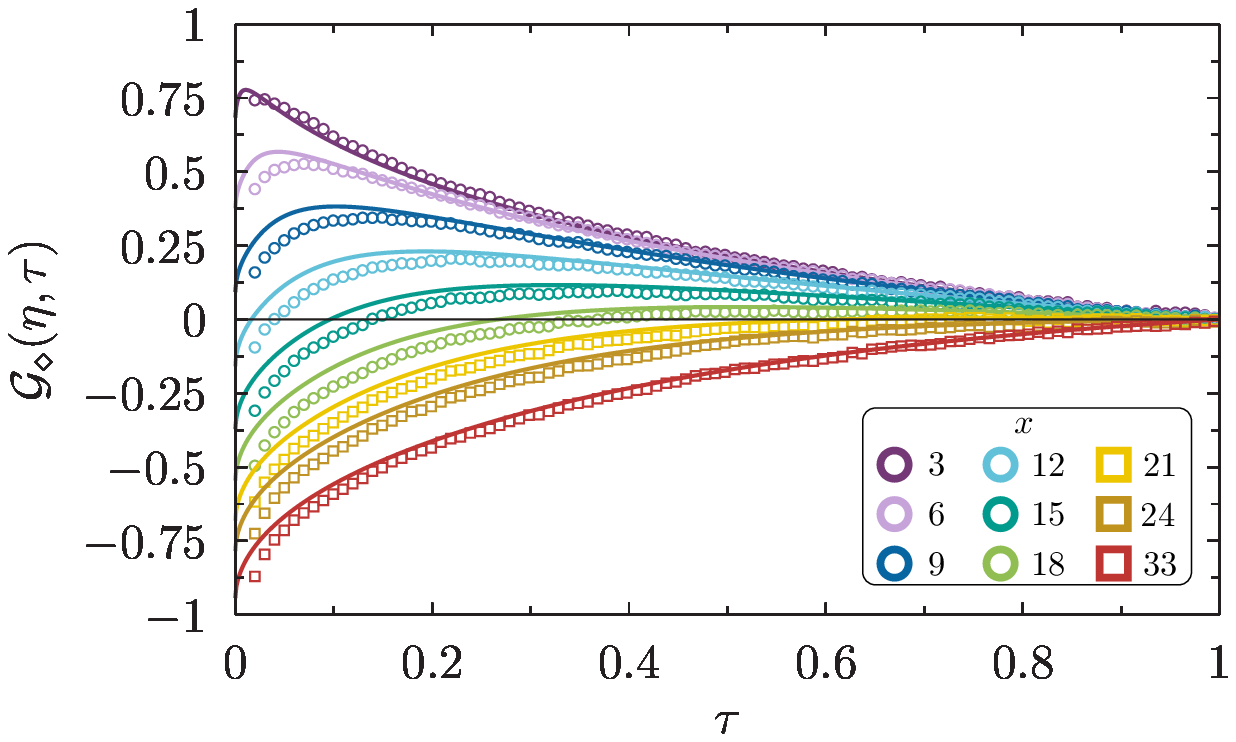}%width=\textwidth
            \caption[]%
            {{\small }}    
            \label{fig_rh_tau}
        \end{subfigure}
%\hspace{30mm}
\vskip\baselineskip
\caption[]
{\small Panel (a) the rhomboidal correlation function $\mathcal{G}_{\diamond}(\eta,\tau)$ as function of $\eta$ for fixed $\tau=2y/R$. Panel (b) $\mathcal{G}_{\diamond}(\eta,\tau)$ as function of $\tau$ for fixed $\eta=x/\lambda$. Data points are obtained from numerical simulations at $T=2$, $R=201$.}
\label{fig_rhomboidal}
\end{figure*}

The next quantity we consider is the correlation function for the rhomboidal pattern. The comparison between theory and numerics is presented in Fig.~\ref{fig_rhomboidal}. In Fig.~\ref{fig_rh_eta}, we plot the correlation function $\mathcal{G}_{\diamond}(\eta,\tau)$ as function of the rescaled coordinate $\eta=x/\lambda$ for several values of $\tau=2y/R$; the vice versa is done in Fig.~\ref{fig_rh_tau}. The analytic result for $\mathcal{G}_{\diamond}(\eta,\tau)$ is provided by (\ref{11032021_1716}) with $\alpha=\sqrt{(1-\tau)/(1+\tau)}$. Focusing firstly on Fig.~\ref{fig_rh_eta}, we observe that for small $\eta$ the correlation function approaches the limiting value
\begin{equation}
\label{ }
\lim_{\eta \rightarrow 0 } \mathcal{G}_{\diamond}(\eta,\tau) = \frac{2}{\pi} \sin^{-1}\left( \alpha^{2} \right) \, ,
\end{equation}
while for $\eta \rightarrow +\infty$ the asymptotic value is the opposite of the one attained for $\eta \rightarrow 0$. The origin of this symmetry can be easily understood by noticing that in both cases the spin fields $\sigma(-x,0)$ and $\sigma(x,0)$ probe either the same phase, or oppositely magnetized phases. Another interesting feature is the rapidity upon which the above limiting values are attained. By taking a first derivative of $\mathcal{G}_{\diamond}(\eta,\tau)$ with respect to $\eta$, we find
\begin{equation}
\label{ }
\partial_{\eta}\mathcal{G}_{\diamond}(\eta,\tau) = \frac{2}{\sqrt{\pi}} \textrm{erf}^{2}(r\eta) \textrm{e}^{-\eta^{2}} \, ,
\end{equation}
meaning that the decay of correlations is exponentially fast along the $x$ axis, in analogy with the rectangular correlation function; see (\ref{18042021_0457}). Such a behavior happens to be in sharp contrast with the algebraic form which characterizes the correlation along the direction parallel to the interface ($y$-direction); see the power law $\propto \sqrt{y/R}$ in (\ref{09042021_1944}). The long-range form of interfacial correlations can be investigated analytically by performing an asymptotic analysis of the function $\mathcal{Y}(|\eta|,r)$ for small $y/R$. This task has been recently carried out in the examination of the three-point correlation function $\langle \sigma(0,y) \sigma(x,0) \sigma(0,-y)\rangle_{-+}$ \cite{ST_threepoint}. In fact, it has been shown in \cite{Squarcini_Multipoint} that the above mentioned three-point correlation function for the Ising model is proportional to the scaling function $\mathcal{Y}(|\eta|,r)$ which occurs in the rhomboidal correlation function. We thus refer the interested reader to Ref. \cite{ST_threepoint} for a detailed mathematical account on this matter.

The remarkable agreement between theory and numerics is confirmed by Fig.~\ref{fig_rh_tau} where the correlation function is plotted as function of $\tau$ for fixed $\eta$. The deviations for small $\tau$ which occur in the plot of Fig.~\ref{fig_rh_tau} are actually expected since that regime is at the boundary of the applicability domain of the analytic result at leading order in powers of $(\xi/R)^{1/2}$. We refer to \cite{DS_twopoint, Squarcini_Multipoint} for a detailed discussion about the theoretical treatment of interface structure corrections and to \cite{DSS1, ST_threepoint} for numerical simulations.

\begin{figure*}[htbp]
\centering
%\hspace{30mm}
        \begin{subfigure}[b]{0.49\textwidth}
            \centering
            \includegraphics[height=45mm]{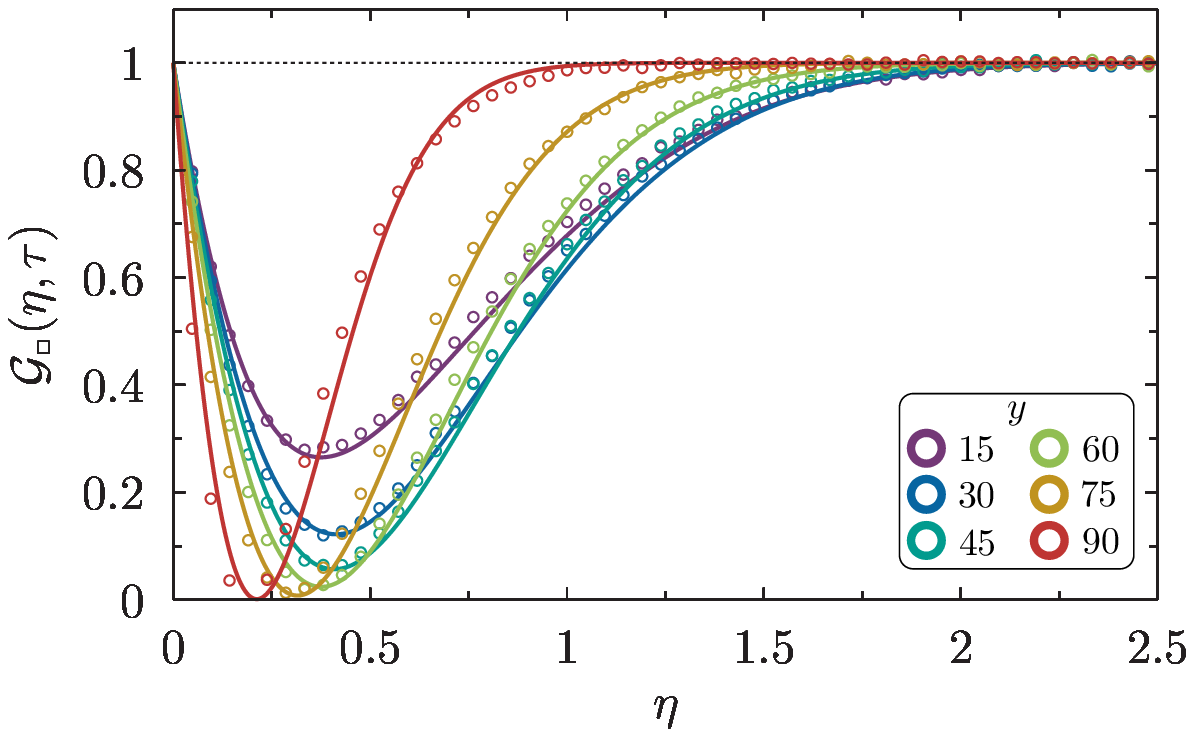}
            \caption[]%
            {{\small }}    
            \label{fig_re_eta}
        \end{subfigure}
\hspace{-3mm}
        \begin{subfigure}[b]{0.490\textwidth}  
            \centering 
            \includegraphics[height=45mm]{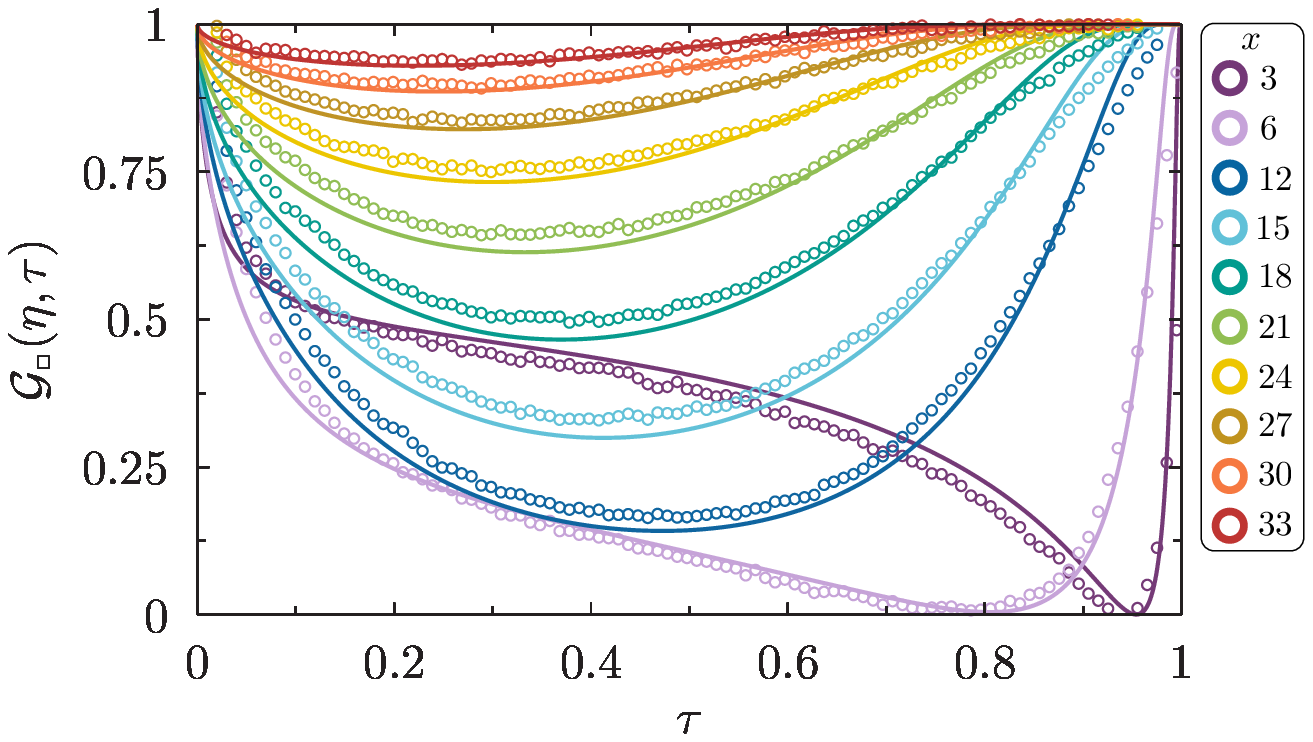}%width=\textwidth
            \caption[]%
            {{\small }}    
            \label{fig_re_tau}
        \end{subfigure}
%\hspace{30mm}
\vskip\baselineskip
\caption[]
{\small Panel (a) the rectangular correlation function $\mathcal{G}_{\oblong}(\eta,\tau)$ as function of $\eta$ for fixed $\tau=2y/R$. Panel (b) $\mathcal{G}_{\oblong}(\eta,\tau)$ as function of $\tau$ for fixed $\eta=x/\lambda$. Data points are obtained from numerical simulations at $T=2$, $R=201$.}
\label{fig_rectangular}
\end{figure*}

Lastly, we discuss the comparison between theory and numerics for spin fields in the rectangular arrangement. The qualitative features discussed for the rhomboidal pattern occur also for the rectangular one. The analytic expression for $\mathcal{G}_{\oblong}(\eta,\tau)$ given in (\ref{11032021_1717}) turns out to be in good agreement with the numerical results shown in Fig.~\ref{fig_rectangular}. The small deviations which are visible in the plot of Fig.~\ref{fig_re_tau} have the same nature of those arising in Fig.~\ref{fig_rh_tau}. 

Although in the field-theoretical approach the bulk correlation length is supposed to be large compared to the lattice spacing $a$, and small compared to the system size, i.e. $a \ll \xi \ll R$, numerical data are found to be in quantitative agreement even when the constraint $\xi \gg a$ is not satisfied. For the temperature $T=2$ considered in the simulations shown in this paper, the correlation length slightly exceeds two lattice spacings. An analogous agreement between theory and simulations for correlation functions within the interfacial region of an Ising interface has been observed for $T=2.15$, corresponding to $\xi \approx 5$ lattice spacings. Furthermore, the striking agreement between theory and numerics even when $\xi$ does not exceeds few lattice spacings has been already observed in recent simulations for the $q$-state Potts model \cite{DSS1}.

%=====================================================================================
\section{Conclusions}
\label{sec4}
In this paper we have illustrated how to find exact closed-form expressions for certain four-point correlation functions for statistical systems exhibiting phase separation on a strip of finite width $R\gg \xi$, with $\xi$ the bulk correlation length. The analytic results derived in this paper are self-consistent with the sole exception of equations (\ref{a004a}), (\ref{a005a}) and (\ref{11032021_1534}), corresponding to the exact probabilistic representation of $n$-point correlation functions which has been recently established from a field-theoretic calculation in Ref. \cite{Squarcini_Multipoint}. We have examined correlation functions with order parameter fields arranged in three configurations: along the interface, in a rhomboidal pattern, and in a rectangular pattern; see Fig.~\ref{fig_ssss}. For each pattern the corresponding correlation function is characterized by two scaling functions whose analytic expression is provided in closed form. These scaling functions are universal in the sense that they are shared by models which exhibit phase separation through a single interface. The exact analytic expressions for the correlation functions provided in this paper allow for a direct examination of general properties of interfacial correlations. In particular, the long-range character of correlations in the direction parallel to the interface and their confinement within the interfacial region follow from a direct inspection of the analytic results we provided. In the second part of the paper we have specialized the analytic results to the Ising model. We have shown that Ising symmetry selects only one of the above mentioned scaling functions (see (\ref{11032021_1715})) and that analytic results are confirmed by high-precision Monte Carlo simulations we performed.

%=====================================================================================
\section*{Acknowledgements}
A. S. is grateful to Gesualdo Delfino for his valuable comments and to Douglas B. Abraham for many interesting discussions and for collaborations on closely related topics.

%=====================================================================================
\appendix

%=====================================================================================
\section{Mathematical details}
\label{Appendix_A}

%=====================================================================================
\subsection{The correlation function $G_{\diamond}$}
The calculation of $G_{\diamond}$ involves certain integrals arising from the cumulative distribution for the passage probability $\Pi_{\diamond}$. In this appendix, we adopt some tricks which enable us to shorten the calculation considerably. Firstly, we write the sharp magnetization profile as follows
\begin{equation}
\label{11032021_2019}
\sigma_{ab \vert i}(x_{i} \vert u_{i}) = \widetilde{\langle\sigma_{i}\rangle} - \frac{1}{2}\Delta\langle\sigma_{i}\rangle s(x_{i}-u_{i}) + \dots \, ,
\end{equation}
where $s(x_{i}-u_{i}) \equiv \textrm{sign}(x_{i}-u_{i})$. Then, we introduce the following shorthand notation for integrals with measure $\Pi_{\diamond}$:
\begin{equation}
\label{31122020_0813}
\llbracket f \rrbracket_{\diamond} = \int_{\mathbb{R}^{4}}\textrm{d}u_{1}\textrm{d}u_{2}\textrm{d}u_{3}\textrm{d}u_{4} \, f(u_{1},u_{2},u_{3},u_{4}) \Pi_{4}^{(\diamond)}(u_{1},u_{2},u_{3},u_{4} \vert \alpha) \, ,
\end{equation}
where $f$ is a function of $u_{1},\dots,u_{4}$ and $\llbracket f \rrbracket_{\diamond}$ stands for its the expectation value. The result of (\ref{31122020_0813}) is a function of $x$ and $y$ but, as we shall see in a while, the dependence through $x$ occurs by means of the rescaled coordinate $\eta=x/\lambda$ while the dependence through $y$ is encoded in $\alpha=\sqrt{(1-\tau)/(1+\tau)}$, with $\tau=2y/R$. Thanks to these notations, the correlation function we are interested in can be written as follows
\begin{equation}
\begin{aligned}
\label{09042021_1652}
G_{\diamond}(x,y) & =  \widetilde{\langle\sigma_{i}\rangle}^{4} - \frac{1}{2}\Delta\langle\sigma_{i}\rangle \widetilde{\langle\sigma_{i}\rangle}^{3} \left( \sum_{i=1}^{4} \llbracket s_{i} \rrbracket_{\diamond} \right) + \frac{1}{4} \left( \Delta\langle\sigma_{i}\rangle \right)^{2} \widetilde{\langle\sigma_{i}\rangle}^{2} \left( \sum_{1 \leqslant i<j \leqslant 4} \llbracket s_{i} s_{j} \rrbracket_{\diamond} \right) \\
& - \frac{1}{8} \left( \Delta\langle\sigma_{i}\rangle \right)^{3} \widetilde{\langle\sigma_{i}\rangle} \left( \sum_{1 \leqslant i<j<k \leqslant 4} \llbracket s_{i} s_{j} s_{k} \rrbracket_{\diamond} \right) + \frac{1}{16} \left( \Delta\langle\sigma_{i}\rangle \right)^{4} \llbracket s_{1} s_{2} s_{3} s_{4} \rrbracket_{\diamond} + \Os(R^{-1/2}) \, ,
\end{aligned}
\end{equation}
where $s_{1} = \textrm{sign}(-u_{1})$, $s_{2} = \textrm{sign}(\sqrt{2}\eta-u_{2})$, $s_{3} = \textrm{sign}(-\sqrt{2}\eta-u_{3})$, and $s_{4} = \textrm{sign}(-u_{4})$. A simple (but tedious) exercise yields the following results:
\begin{equation}
\begin{aligned}
\label{09042021_1653}
\sum_{i=1}^{4} \llbracket s_{i} \rrbracket_{\diamond} & = 0 \, , \\
\sum_{1 \leqslant i<j \leqslant 4} \llbracket s_{i} s_{j} \rrbracket_{\diamond} & = 1 - 2 \textrm{erf}(|\eta|) + \frac{2}{\pi} \sin^{-1}\left( \alpha^{2} \right) + 16 T(\sqrt{2}\eta,r) \equiv \mathcal{C}_{\diamond}(\eta,\tau) \, , \\
\sum_{1 \leqslant i<j<k \leqslant 4} \llbracket s_{i} s_{j} s_{k} \rrbracket_{\diamond} & = 0 \, , \\
\llbracket s_{1} s_{2} s_{3} s_{4} \rrbracket_{\diamond} & = \frac{2}{\pi} \sin^{-1}\left( \alpha^{2} \right) - 2 \mathcal{Y}(|\eta|,r) \equiv \mathcal{G}_{\diamond}(\eta,\tau) \, ,
\end{aligned}
\end{equation}
with $\alpha$ and $r$ given by (\ref{12042021_0913}) and (\ref{12042021_0914}), respectively, $T$ is Owen's function \cite{Owen1956, Owen1980}
\begin{equation}
\label{31122020_0850}
T(h,a) = \frac{1}{2\pi} \int_{0}^{a} \textrm{d}x \, \frac{ \textrm{e}^{-h^{2}\frac{1+x^{2}}{2}} }{ 1+x^{2} } \, ,
\end{equation}
and $\mathcal{Y}$ is the function defined by (\ref{31122020_0849}). It is useful to recall some basic properties satisfied by Owen's $T$ function \cite{Owen1980}. The function (\ref{31122020_0850}) satisfies $T(\pm \infty,a)=T(h,0)=0$, the symmetry $T(h,a)=T(-h,a)$, and it reduces to elementary functions for special values of its arguments; for instance:
\begin{equation}
\label{31122020_0855}
T(0,a) = \frac{1}{2\pi} \tan^{-1}(a) \, .
\end{equation}
Other useful properties are the integral representation
\begin{equation}
\label{23042021_1457}
T(\sqrt{2}\eta,r) = -\frac{1}{2\sqrt{\pi}} \int_{-\infty}^{\eta}\textrm{d}u \, \textrm{e}^{-u^{2}} \textrm{erf}(ru) \, ,
\end{equation}
and the functional identity \cite{Owen1956}
\begin{equation}
\begin{aligned}
\label{09042021_1657}
T(h,a) + T(ah,1/a) & = \frac{1}{2}g(h) + \frac{1}{2}g(ah) - g(h)g(ah) \, , \qquad a\geqslant0 \, \\
g(x) & = \frac{1+\textrm{erf}(x/\sqrt{2})}{2} \, .
\end{aligned}
\end{equation}
Equations (\ref{31122020_0843}) and (\ref{11032021_1716}) for the rhomboidal correlation function follow directly from (\ref{09042021_1652}) and (\ref{09042021_1653}).

%=====================================================================================
\subsection{The correlation function $G_{\oblong}$}
Let us consider now the rectangular correlation function. Since for this configuration $y_{1}=y_{2}$ and $y_{3}=y_{4}$ the corresponding correlation coefficients degenerate to unity, i.e., $\rho_{12}=1$ and $\rho_{34}=1$, respectively. The corresponding passage probability is found by taking the double limit $\rho_{12} \rightarrow1$ and $\rho_{34} \rightarrow1$, the latter yields
\begin{equation}
\begin{aligned}
\label{11032021_1229}
\Pi_{4}^{(\oblong)}(u_{1},u_{2},u_{3},u_{4} \vert \rho_{23}) & = \lim_{\rho_{12} \rightarrow 1} \lim_{\rho_{34} \rightarrow 1} \Pi_{4}(u_{1},u_{2},u_{3},u_{4} \vert \textsf{R}_{1234}) \\
& = \delta(u_{1}-u_{2}) \delta(u_{3}-u_{4}) \Pi_{2}(u_{1},u_{4} \vert \rho_{23}) \, ,
\end{aligned}
\end{equation}
where $\Pi_{2}(u_{1},u_{4} \vert \rho_{23})$ is a bivariate normal distribution with correlation coefficient $\rho_{23}$. By inserting (\ref{11032021_1229}) into the probabilistic representation for the correlation function, we find
\begin{equation}
\begin{aligned}
\label{09042021_1201}
G_{\oblong}(x,y) & = \int_{\mathbb{R}^{4}}\textrm{d}u_{1}\textrm{d}u_{2}\textrm{d}u_{3}\textrm{d}u_{4} \Bigl[ \widetilde{\langle\sigma_{i}\rangle} - \frac{\Delta\langle\sigma_{i}\rangle}{2} \textrm{sign}(\sqrt{2}\chi-u_{1}) \Bigr]  \Bigl[ \widetilde{\langle\sigma_{i}\rangle} - \frac{\Delta\langle\sigma_{i}\rangle}{2} \textrm{sign}(-\sqrt{2}\chi-u_{2}) \Bigr] \\
& \times \Bigl[ \widetilde{\langle\sigma_{i}\rangle} - \frac{\Delta\langle\sigma_{i}\rangle}{2} \textrm{sign}(-\sqrt{2}\chi-u_{3}) \Bigr] \Bigl[ \widetilde{\langle\sigma_{i}\rangle} - \frac{\Delta\langle\sigma_{i}\rangle}{2} \textrm{sign}(\sqrt{2}\chi-u_{4}) \Bigr] \Pi_{4}^{(\oblong)}(u_{1},u_{2},u_{3},u_{4} \vert \rho_{23}) \\
& + \Os(R^{-1/2}) \, .
\end{aligned}
\end{equation}
A rearrangement of the terms in (\ref{09042021_1201}) gives
\begin{equation}
\begin{aligned}
\label{11032021_1304}
G_{\oblong}(x,y) & =  \widetilde{\langle\sigma_{i}\rangle}^{4} - \frac{1}{2}\Delta\langle\sigma_{i}\rangle \widetilde{\langle\sigma_{i}\rangle}^{3} \left( \sum_{i=1}^{4} \llbracket s_{i} \rrbracket_{\oblong}^{\vphantom{\oblong}} \right) + \frac{1}{4} \left( \Delta\langle\sigma_{i}\rangle \right)^{2} \widetilde{\langle\sigma_{i}\rangle}^{2} \left( \sum_{1 \leqslant i<j \leqslant 4} \llbracket s_{i} s_{j} \rrbracket_{\oblong}^{\vphantom{\oblong}} \right) \\
& - \frac{1}{8} \left( \Delta\langle\sigma_{i}\rangle \right)^{3} \widetilde{\langle\sigma_{i}\rangle} \left( \sum_{1 \leqslant i<j<k \leqslant 4} \llbracket s_{i} s_{j} s_{k} \rrbracket_{\oblong}^{\vphantom{\oblong}} \right) + \frac{1}{16} \left( \Delta\langle\sigma_{i}\rangle \right)^{4} \llbracket s_{1} s_{2} s_{3} s_{4} \rrbracket_{\oblong}^{\vphantom{\oblong}} + \Os(R^{-1/2}) \, ,
\end{aligned}
\end{equation}
with $s_{1} = \textrm{sign}(\sqrt{2}\chi-u_{1})$, $s_{2} = \textrm{sign}(-\sqrt{2}\chi-u_{2})$, $s_{3} = \textrm{sign}(-\sqrt{2}\chi-u_{3})$, $s_{4} = \textrm{sign}(\sqrt{2}\chi-u_{4})$, and the compact notation
\begin{equation}
\label{12033021_1151}
\llbracket f \rrbracket_{\oblong}^{\vphantom{\oblong}} = \int_{\mathbb{R}^{4}}\textrm{d}u_{1}\textrm{d}u_{2}\textrm{d}u_{3}\textrm{d}u_{4} \, f(u_{1},u_{2},u_{3},u_{4}) \Pi_{4}^{(\oblong)}(u_{1},u_{2},u_{3},u_{4} \vert \rho_{23})
\end{equation}
has been adopted. By using the integral representation (\ref{23042021_1457}) of Owen's $T$ function, a simple (but rather long) calculation entails
\begin{equation}
\begin{aligned}
\sum_{i=1}^{4} \llbracket s_{i} \rrbracket_{\oblong}^{\vphantom{\oblong}} & = 0 \, , \\
\sum_{1 \leqslant i<j \leqslant 4} \llbracket s_{i} s_{j} \rrbracket_{\oblong}^{\vphantom{\oblong}} & = 2  - 4 \textrm{erf}(|\chi|) - 16 T(\sqrt{2}\chi,\sqrt{\tau}) + 16 T(\sqrt{2}\chi,1/\sqrt{\tau}) \equiv \mathcal{C}_{\oblong}(\eta,\tau) \, , \\
\sum_{1 \leqslant i<j<k \leqslant 4} \llbracket s_{i} s_{j} s_{k} \rrbracket_{\oblong}^{\vphantom{\oblong}} & = 0 \, , \\
\llbracket s_{1} s_{2} s_{3} s_{4} \rrbracket_{\oblong}^{\vphantom{\oblong}} & =  5 - 4 \textrm{erf}(|\chi|) - 16 T(\sqrt{2}\chi,\sqrt{\tau}) - 16 T(\sqrt{2}\chi,1/\sqrt{\tau}) \equiv \mathcal{G}_{\oblong}(\eta,\tau) \, .
\end{aligned}
\end{equation}
The results (\ref{11032021_1305}) and (\ref{11032021_1717}) given in the main body of the paper follow straightforwardly.

\subsection{A useful identity}
Here, we prove the following identity
\begin{equation}
\begin{aligned}
\label{idA13}
f(x) & = \frac{4}{\pi^{2}} \left( \sin^{-1}x \right)^{2} + \frac{4}{\pi^{2}} \int_{0}^{x^{2}} \frac{\textrm{d}u}{\sqrt{1-u^{2}}} \sin^{-1}\left( u \frac{1-x^{2}}{x^{2}-u^{2}} \right) \, \\
& = \frac{2}{\pi} \sin^{-1}\left( x^{2} \right) \, ,
\end{aligned}
\end{equation}
with $0 \leqslant x \leqslant 1$. By taking the first derivative with respect to $x$, we obtain
\begin{equation}
\begin{aligned}
\label{}
f^{\prime}(x) & = \frac{8}{\pi^{2}} \frac{\sin^{-1}x}{\sqrt{1-x^{2}}} + \frac{4}{\pi} \frac{x}{\sqrt{1-x^{4}}} - \frac{8x}{\pi^{2}} \int_{0}^{x^{2}} \textrm{d}u \, \frac{u}{(x^{2}-u^{2})\sqrt{x^{4}-u^{2}}} \, ,
\end{aligned}
\end{equation}
the integral in the third term reads
\begin{equation}
\label{ }
\int_{0}^{x^{2}} \textrm{d}u \, \frac{u}{(x^{2}-u^{2})\sqrt{x^{4}-u^{2}}} = \frac{\sin^{-1}x}{x\sqrt{1-x^{2}}} \, ,
\end{equation}
thus, it follows that
\begin{equation}
\begin{aligned}
\label{}
f^{\prime}(x) & = \frac{4}{\pi} \frac{x}{\sqrt{1-x^{4}}} \, .
\end{aligned}
\end{equation}
The identity (\ref{idA13}) is established by integrating back with respect to $x$ and fixing the integration constant by using the boundary condition, $f(0)=0$, or $f(1)=1$.

%==================================================================================
\bibliographystyle{unsrt}
\bibliography{bibliography}{}

%==================================================================================
\end{document}